\documentstyle[preprint,eqsecnum,prd,aps]{revtex}
\begin{document} 

\preprint{
\vbox{
\halign{&##\hfil\cr
	& ANL-HEP-PR-96-60 \cr
	& November 22, 1996 \cr}}
}
\title{Isolated Photons at Hadron Colliders at $O(\alpha\alpha_s^2)$
(I): Spin Averaged Case}

\author{L. E. Gordon}
\address{High Energy Physics Division, Argonne National Laboratory,
	Argonne, IL 60439}
\maketitle
\begin{abstract} 

The cross sections for isolated and non-isolated prompt photon production with 
unpolarized hadron beams are studied at order $\alpha\alpha_s^2
$. Two methods of performing the calculations are compared. One uses
purely analytic techniques and the second uses a combination of analytic and 
Monte Carlo techniques to perform the phase-space integrations. The
results of the  analytic and Monte Carlo methods are compared both
before and after isolation cuts are placed on the photon. Fragmentation
contributions are included at next-to-leading order in the analytic
case, and the question of the infrared sensitivity of this contribution
once isolation is implemented is addressed.
Numerical results are presented and compared with the latest CDF 
data for $p\bar{p}\rightarrow \gamma+X$ using the latest parton densities for 
the proton. 
\vspace{0.2in}
\end{abstract}
\narrowtext

\section{Introduction}
Prompt photon production is now firmly established as one of the most
powerful tools for uncovering information on the structure of
hadrons and is therefore one of the best ways known for testing
perturbative quantum chromodynamics (QCD) \cite{owens}. This is due in
large part to the fact that photons couple directly to quarks via the
well understood point-like electromagnetic interaction, and therefore
provide an incisive probe of the short distance dynamics of the
hard scattering process involving the strong interaction. Since the cross 
section for prompt photon
production is dominated by the subprocess $qg\rightarrow\gamma q$
in hadronic collisions already in leading order (LO), it has proved very 
useful for providing information on the gluon distributions, $g(x,Q^2)$,
of hadrons at fixed target facilities. At colliders the situation is
more complicated due to the necessity of imposing isolation restrictions
on the photon in order to identify the signal from among the copious hadronic
debris. This restricts the usefulness of the process for extracting
information on gluon distributions since isolation is not yet
fully understood theoretically and there is some controversy pertaining
to the use of the conventional factorization theorem in the infrared
regions of phase space for the fragmentation contribution in NLO 
\cite{bergerqui1,auretal} when isolation cuts are imposed. These
difficulties are very unwelcome since prompt photon production probes
the parton densities of the proton in the region $x\sim
x_T=2p^{\gamma}_T/\sqrt{s}=10^{-2}$ for $p_T^{\gamma}=10$ GeV at
pseudo-rapidity, $\eta=0$ at the Fermilab TEVATRON. Information gained
about the gluon densities at colliders could thus be very important for
filling in the gaps left by deep inelastic processes which do not
constrain it very tightly.   

Unfortunately the comparisons of the theoretical predictions and the
collider data have proven to be very problematic. For example, the
original comparisons between theoretical predictions and the data from
the CDF collaboration \cite{abe} showed differences as large as a factor
of $1.5$ or $2$ in the low $p_T^{\gamma}$ region, the data always being
larger. Since that time things have improved, mainly due to improvements
in our knowledge of the parton densities, but significant differences
still remained which were made even more problematic by uncertainties in
the theoretical calculations. Despite this, significant progress has also
been made on the theoretical front as will be demonstrated in this
paper. 

Not taking into account calculations done in LO, there have been three
previous independent calculations of the isolated prompt photon cross
section in NLO. The first attempt was made by Aurenche et al. \cite{auretal1}. 
In this calculation isolation was implemented approximately from a modification
of the inclusive cross section
calculated in \cite{auretal2}, which was one of the earliest complete NLO
calculation ever performed in perturbative QCD. Photon isolation was achieved 
by modifying 
the large logarithms which occur in the higher order contributions to
the non-fragmentation processes when the photon is produced collinearly to a 
quark. The fragmentation contributions were calculated in LO and a cut
was placed on the remnants of the fragmenting parton, and the
fragmentation scale was also modified accordingly. This method,
although approximate, retained the most important terms for the
isolation procedure, such as logarithms of the half-angle of the isolation 
cone $\ln\delta$ and thus gave quite accurate results for the
isolated cross section. Terms such as $\ln\epsilon$, where $\epsilon$ is
the energy resolution parameter (introduced in section II) were not all
retained, hence the procedure would not give accurate results for very small 
values of this parameter, but in general, dependence of the cross section on 
this parameter is not very large. Thus the only major area where the
calculation was deficient was in the use of LO matrix elements for the
fragmentation contributions and the use of LO asymptotic fragmentation
functions, both of which were all that were available at the time. 

The next major development was the calculation of Baer et al.
\cite{baeretal} which dealt with isolation exactly in NLO for the
non-fragmentation contributions by the use of a
combination of Monte Carlo and analytic techniques to do the phase space
integrals. This method is outlined in section IIB(3). The only drawback of
this calculation was again that the fragmentation contributions were
included in LO and LO asymptotic fragmentation functions were
used. 

Important theoretical work was also done by Berger and Qiu who studied the 
isolated cross section in NLO \cite{bergerqiu4} in some detail. 
They studied the dependence of the cross section on the isolation parameters 
and pointed out that the direct 
contribution should have a $\ln\epsilon$ dependence by examining the general 
structure of the isolated cross section in the soft limit. They suggested
that the isolated cross section may be conveniently written as the inclusive 
cross section minus a subtraction piece which removes hadronic energy from 
around the photon whenever this energy is too large to pass the experimental 
cuts. Their work helped to clarify many of the theoretical issues and 
highlighted some of the problems involved in defining the isolated cross 
section. Their numerical results were based on the two previous calculations 
of refs.\cite{auretal1,baeretal}.

The next independent calculation came with the calculation of Gordon and 
Vogelsang \cite{gorvogel3} which was confronted with the CDF data in a paper by
Gl\"{u}ck et al. \cite{reya}. In this paper the isolated cross section was
calculated using an approximate isolation method which was developed in 
\cite{gorvogel3} from a modification of the inclusive cross sections
calculated in \cite{gorvogel} for the direct part and in \cite{aversa}
for the fragmentation part. In \cite{gorvogel3} the basic idea of
writing the isolated cross section as the inclusive cross section minus
a subtraction piece, suggested by Berger and Qiu \cite{bergerqiu4} was 
implemented and worked out in detail at the partonic level for both the direct 
and fragmentation contributions. The most significant improvements made were 
thus inclusion of the fragmentation contributions in NLO and the use of parton 
to photon fragmentation functions evolved in NLO QCD \cite{vogtfrag}.
The isolation method used, although, strictly speaking, should be valid only 
in the limit of a very small isolation cone, just as was the case for that of 
Aurenche et al., \cite{auretal1}, was shown to reproduce many of the
features of the exact calculation \cite{gorvogel3} as well as to improve the
agreement between theory and experiment when applied at the Fermilab TEVATRON
with the inclusion of fragmentation contributions at NLO \cite{reya}.
One significant difference between the calculations of Gl\"{u}ck et al. 
\cite{gorvogel3,reya} and Aurenche et al. \cite{auretal1} was that terms such 
$\ln \epsilon$ were retained in the former, leading to a better behaviour 
when $\epsilon$ is varied.  

In \cite{gorvogel3} we tested the accuracy of our analytical isolation method 
for various values of the
isolation parameters by comparing results to our own version of a Monte Carlo
method. In that case the Monte Carlo calculation was obtained from the
fully analytic results by subtracting a piece from the latter that involved
the isolation cuts. The subtraction piece was integrated over the restricted 
phase-space inside the isolation cone by Monte Carlo integration (I refer to
\cite{gorvogel3} for the exact details, but provide an outline in the
next section). Our Monte Carlo calculation was thus not entirely independent of
our analytic one. The usual combined analytic/Monte Carlo method used in 
\cite{baeretal} is different as outlined below, and is in many
ways more versatile but should in the end give the same results. 

Since our results for the isolated cross section were first 
published there have been attempts, mainly by experimentalists, to compare our 
results with that of Baer et al. \cite{baeretal} to see if the predictions for 
the isolated cross section differential in $p^{\gamma}_T$ are indeed steeper in 
the low $p^{\gamma}_T$ region after 
inclusion of the NLO fragmentation contributions, thereby accounting for some 
of the discrepancy in slope between theoretical predictions and experimental 
data in this region \cite{experiment}. Our analysis \cite{gorvogel3,reya} showed
clearly that, as one would expect, 
inclusion of the fragmentation contributions in NLO does indeed give a steeper
cross section in this region. Surprisingly a comparison of our numerical results
with that from the program of Baer et al.\cite{baeretal} showed that we only 
had a constant
$O(13\%)$ enhancement at $all$ $p^{\gamma}_T$ values \cite{kuhlmann}. This 
result was very puzzling as no one can dispute that the NLO fragmentation 
contributions do have a steeper $p^{\gamma}_T$ dependence than the
corresponding LO ones as demonstrated in \cite{reya}. The source of this
result was thus far not determined as our calculation methods were 
different and the only thing common in both cases is that both results were
compared to the analytic calculation of Aurenche et al. in the non-isolated 
case \cite{auretal1}, and agreement was found. 
In our case a detailed term-by-term algebraic comparison was also made. 

In principle the analytic/Monte Carlo method of Baer et al. should give exactly
the same results for the direct contributions to the inclusive prompt photon 
cross section, before isolation cuts are implemented, as the calculations of 
Aurenche et al. \cite{auretal1} and Gordon and Vogelsang \cite{gorvogel} 
to within the errors of the Monte Carlo 
integration if the same parameters are used in the calculations. Once isolation
cuts are implemented one may possibly expect the exact Monte Carlo method to
give the more accurate results, but since the approximate methods of
refs.\cite{auretal2} and \cite{gorvogel3} retained the most important
terms then any differences should be very small. The results of the
calculations of Aurenche et al. refs.\cite{auretal1} and Gordon and
Vogelsang \cite{gorvogel} agreed for the
non-isolated case, and thus any discrepancy between the results of the the 
calculations of Baer et al. \cite{baeretal} and the isolated cross
section \cite{gorvogel3,reya} provides a strong
motivation for repeating the calculation of the former and making a more
detailed comparison.   
The main motivation for repeating the Monte Carlo calculation was thus
to confirm our conclusions in \cite{gorvogel3} and \cite{reya} and hopefully 
clear up some of the confusion by finding the source of the discrepancy between
our results and that of Baer et al. \cite{baeretal}. Details of the
Monte Carlo calculation are repeated in ref.\cite{gordon} where the polarized 
cross section is also calculated using the same method. A new feature 
in the present calculation as compared to Baer et al. \cite{baeretal} is that 
the various subprocess contributions are kept
separate which allows one to tell how much of the cross section is due
to $qg$ or $q\bar{q}$ scattering, for example as well as allowing a
process by process comparison with the analytic calculation where they
were kept separate \cite{gorvogel}. This information could be
useful if one is interested in the sensitivity of the cross section to
the gluon densities. For the polarized case, the present calculation is 
the first using the analytic/Monte Carlo method. 

In this paper I therefore compare again the analytic results with the Monte 
Carlo results, where in this case, Monte Carlo means the full
independent analytic/Monte Carlo method. It will be shown that the two
methods agree extremely well both before and after isolation cuts are
implemented. It turns out that the discrepancy between the results of
Baer et al. \cite{baeretal} and Gl\"{u}ck et al. \cite{reya} was 
due to an error in the 
version of the code of the former used by the experimentalists. This 
error caused the non-fragmentation part of the predicted cross section to 
have a steeper $p^{\gamma}_T$ dependence than those of both Aurenche et
al. \cite{auretal2} and \cite{gorvogel3,reya} even before isolation cuts are 
imposed. It
also had the unfortunate effect of cancelling out the increase in slope due to
the inclusion of the NLO fragmentation contributions after isolation,
given the choice of cut-off parameters used. Another choice of cut-off
parameters would likely have given very different results.

All calculations now agree fully for the direct contribution, and it is 
therefore now possible 
to study whether the use of modern parton distributions coupled with the 
inclusion of the NLO fragmentation contributions will improve the agreement
between theory and experiment. This will have a direct bearing on recent
studies \cite{reno} which suggested that initial state gluon radiation
can be used to explain the difference in slope between theory and
data. A reduction of the discrepancy would suggest that less initial
state radiation would be needed.

The rest of this paper is as follows; In section II a brief
theoretical background to the calculations is given including an outline of 
the analytic/Monte
Carlo method used, as well as the approximate analytic method of
isolation. In section III a fairly detailed comparison of the
analytic and Monte Carlo methods is made and comparisons with the CDF
data is presented, and in section IV the conclusions are given.  

\section{Isolated Prompt Photons}

In this section I briefly outline the calculational techniques used to
obtain the inclusive and isolated prompt photon cross sections in both
the analytic and Monte Carlo cases. 

Contributions to the prompt photon cross section are usually separated
into two classes in both LO and NLO. There are the so-called direct
processes, $ab\rightarrow \gamma c$ in LO and $ab\rightarrow \gamma c
d$, in NLO, $a,b,c$ and $d$ referring to partons, where the photon is produced 
directly in the hard scattering. In addition there are the fragmentation
contributions where the photon is produced via bremsstrahlung off a
final state quark or gluon, $ab\rightarrow c d (e)$ followed by
$c\rightarrow \gamma + X$ for instance. 

Experimentally, a prompt photon is considered isolated if inside a cone
of radius $R$ centered on the photon the hadronic energy is less than 
$\epsilon E_\gamma$, where $E_\gamma$ is the photon energy and
$\epsilon$ is the energy resolution parameter, typically $\epsilon \sim 0.1$. 
The radius of the circle defined by the isolation cone is given in the 
pseudo-rapidity $\eta$ and azimuthal angle $\phi$-plane by
$R=\sqrt{(\Delta \eta)^2 + (\Delta \phi)^2}$. In the case of a small
cone the parameter used is the half angle of the cone, $\delta$, where
$\delta \approx R$ for small rapidities of the photon. The exact
relation is $R=\delta/{\cosh \eta}$.

\subsection{The LO Case}

In LO, $O(\alpha \alpha_s)$, the direct subprocesses contributing 
to the cross section are
\begin{eqnarray}
qg&\rightarrow& \gamma q \nonumber \\
q\bar{q}&\rightarrow& \gamma g.
\end{eqnarray}
In addition there are the fragmentation processes
\begin{eqnarray}
qg &\rightarrow& q g \nonumber \\
qq &\rightarrow& q q \nonumber \\
qq' &\rightarrow& q q' \nonumber \\
q\bar{q} &\rightarrow& q \bar{q} \nonumber \\
qg &\rightarrow& q g \nonumber \\
q\bar{q} &\rightarrow& g g \nonumber \\
gg &\rightarrow& g g \nonumber \\
gg &\rightarrow& q \bar{q} 
\end{eqnarray}
where one of the final state partons fragments to produce the photon,
i.e., $q (g)\rightarrow \gamma + X$.

In the direct processes in LO, the photon is always isolated since it
must always balance the transverse momentum $p_T$ of the other final
state parton and is thus always in the opposite hemisphere. In this case 
the differential cross section is given by
\begin{equation}
E_\gamma\frac{d\sigma_{dir}^{LO}}{d^3p_\gamma}=\frac{1}{\pi S}\sum_{i,j}
\int^V_{V
W}\frac{dv}{1-v}\int^1_{VW/v}f^a_1(x_1,M^2)f^b_2(x_2,M^2)\frac{1}{v}\frac
{d\hat{\sigma}_{ab\rightarrow\gamma}}{dv}\delta(1-w)
\end{equation}
where $S=(P_1+P_2)^2$, $V=1+T/S$, $W=-U/(T+S)$, $v=1+\hat{t}/\hat{s}$,
$w=-\hat{u}/(\hat{t}+\hat{s})$, $\hat{s}=x_1 x_2 S$, and
$T=(P_1-P_\gamma)^2$ and $U=(P_2-P_\gamma)^2$. As usual the Mandelstam
variables are defined in the upper case for the hadron-hadron system
and in lower case in the parton-parton system. $P_1$ and $P_2$ are the 
momenta of the incoming hadrons and $f^a_1(x_1,M^2)$ and
$f^b_2(x_2,M^2)$ represent the respective probabilities of finding parton $i$ 
and $j$ in hadrons $1$ and $2$ with momentum fractions $x_1$ and $x_2$
at scale $M^2$. 
 
For the fragmentation processes, the photon is always produced nearly
collinearly to the fragmenting parton and an isolation cut must be
placed on the cross section to remove the remnants of the fragmenting
parton if it has more energy than $\epsilon E_{\gamma}$. In this case this 
restriction is quite easy to implement. The inclusive differential cross 
section is given by
\begin{eqnarray}
E_\gamma\frac{d\sigma_{frag}^{incl}}{d^3p_\gamma}&=&\frac{1}{\pi
S}\sum_{a,b,c}
\int^1_{1-V+VW}\frac{dz}{z^2}\int^{1-(1-V)/z}_{V
W/z}\frac{dv}{1-v}\int^1_{VW/vz}\frac{dw}{w}f^a_1(x_1,M^2)f^b_2(x_2,M^2)
\nonumber \\
&\times &\frac{1}{v}
\frac{d\hat{\sigma}_{ab\rightarrow c}}{dv}\delta(1-w)D^\gamma_c(z,M_f^2),
\end{eqnarray}
where $D_{\gamma/c}(z,M_f^2)$ represents the probability that the parton
labelled $c$ fragments to a photon with a momentum fraction $z$ of its
own momentum at scale $M_f^2$. This is the non-perturbative
fragmentation function which must be obtained from experiment at some
scale and evolved to $M_f^2$ using the usual evolution equations.  
This means that in order to obtain the isolated cross section we
simply have to cut on the variable $z$. If isolation is defined in the
usual way by only accepting events with hadronic energy less than
fraction $\epsilon$ in a cone of radius $R=\sqrt{(\Delta
\phi)^2+(\Delta\eta)^2}$ drawn in the pseudo-rapidity azimuthal angle
plane around the photon, then the hadronic remnants of the fragmenting
parton will always automatically be inside the cone with the photon, for
suitable choices of $M_f$, and the isolated cross section is given by the 
equation
\begin{eqnarray}
E_\gamma\frac{d\sigma_{frag}^{isol}}{d^3p_\gamma}&=&\frac{1}{\pi
S}\sum_{a,b,c}
\int^1_{MAX[z_{min},1/(1+\epsilon)]}\frac{dz}{z^2}\int^{1-(1-V)/z}_{V
W/z}\frac{dv}{1-v}\int^1_{VW/vz}\frac{dw}{w}f^a_1(x_1,M^2)f^b_2(x_2,M^2)
\nonumber \\ &\times &\frac{1}{v}\frac
{d\hat{\sigma}_{ab\rightarrow c}}{dv}\delta(1-w)D^\gamma_c(z,M_f^2),
\end{eqnarray}
where $z_{min}=1-V+VW$. 
It is also suggested that the fragmentation scale should be replaced by
$(R M_f)$ or $(\delta M_f)$ in order to ensure that all fragmentation remnants 
are radiated
inside the cone \cite{bergerqiu4}, but this argument not universally
accepted. It was shown in \cite{gorvogel3} that the choice is
numerically irrelevant, since the dependence of the cross section on the
fragmentation scale is negligible after isolation in NLO.       

\subsection{The NLO Case}

\subsubsection{The Non-Fragmentation Contribution}

At $O(\alpha\alpha_s^2)$ there are virtual corrections 
to the LO non-fragmentation processes of eq.(2.1), as well as the further 
three-body processes:
\begin{mathletters}\label{eq:1}
\begin{eqnarray}
g &+q  \rightarrow g + q + \gamma\label{eq:11}\\
g &+ g \rightarrow q +\bar{q} + \gamma\label{eq:12}\\
q &+ \bar{q} \rightarrow g + g + \gamma\label{eq:13}\\
q &+ q \rightarrow  q + q + \gamma\label{eq:14}\\
\bar{q} &+ q \rightarrow \bar{q} + q + \gamma\label{eq:15}\\
q &+ \bar{q} \rightarrow q' + \bar{q}' + \gamma\label{eq:16}\\
q &+ q' \rightarrow q + q' + \gamma\label{eq:17}
\end{eqnarray}
\end{mathletters} 

In principle the fragmentation processes of eq.(2.2) should now be
calculated to $O(\alpha_s^3)$ and convoluted with the NLO photon
fragmentation functions whose leading behaviour is $O(\alpha/\alpha_s)$
leading to contributions of $O(\alpha\alpha_s^2)$. For the inclusive
cross section fragmentation processes can contribute $50-60\%$ of
the cross section at the lowest $p_T^{\gamma}$ values at TEVATRON
energies. Numerically the fragmentation processes are not as significant 
after isolation cuts are implemented, but for a theoretically consistent
calculation they should nevertheless be included as they help to reduce
scale dependences, and as was demonstrated in ref.\cite{reya} they also
help to improve the agreement between theory and experiment in the low
$p_T^\gamma$ region.

The direct contribution to the inclusive cross section is given by
\begin{eqnarray}
E_\gamma\frac{d\sigma^{incl}_{dir}}{d^3p_\gamma}&=&\frac{1}{\pi S}
\sum_{a,b}\int^V_{V
W}\frac{dv}{1-v}\int^1_{VW/v}\frac{dw}{w}f^a_1(x_1,M^2)f^b_2(x_2,M^2)
\nonumber \\ &\times&\left[ \frac{1}{v}\frac
{d\hat{\sigma}_{ab\rightarrow\gamma}}{dv}\delta(1-w)+
\frac{\alpha_s(\mu^2)}{2\pi}
K_{ab\rightarrow\gamma}(\hat{s},v,w,\mu^2,M^2,M_f^2)\right],
\end{eqnarray}
where $K_{ab\rightarrow\gamma}(\hat{s},v,w,\mu^2,M^2,M_f^2)$ represents
the higher corrections to the hard subprocess cross sections calculated
in \cite{gorvogel} and $\mu$ is the renormalization scale. 

In \cite{gorvogel3} the isolated cross section is written as the
inclusive cross section minus a subtraction piece as first suggested in
\cite{bergerqiu4}
\begin{equation}
E_\gamma\frac{d\sigma^{isol}_{dir}}{d^3p_\gamma}=
E_\gamma\frac{d\sigma^{incl}_{dir}}{d^3p_\gamma}-
E_\gamma\frac{d\sigma^{sub}_{dir}}{d^3p_\gamma},
\end{equation}
$E_\gamma\frac{d\sigma^{sub}_{dir}}{d^3p_\gamma}$ being the cross section
for producing a prompt photon with energy $E_\gamma$ which is
accompanied by more hadronic energy than $\epsilon E_\gamma$ inside the
cone. The question is then how to calculate the subtraction piece. In
\cite{gorvogel3} it is calculated by an approximate analytic method for
a small cone of half angle $\delta$ as well as for a cone of radius $R$
as defined above using Monte Carlo integration methods. The complete
details of the calculation can be found in ref.\cite{gorvogel3}. The
final form for the subtraction piece assuming a small cone of half angle
$\delta$ is given by
\begin{equation}
E_\gamma\frac{d^3\sigma^{sub}_{dir}}{d^3p_\gamma}= A\ln \delta +B
+C\delta^2\ln\epsilon,
\end{equation}
where $A,B$ and $C$ are functions of the kinematic variables of the
photon and $\epsilon$. The functions $A$ $B$ and $C$ were calculated in
ref.\cite{gorvogel3} and can be found in the appendix of that paper. As
noted earlier there is a term proportional $\ln\epsilon$ as anticipated
in ref.\cite{bergerqiu4} which was not retained in ref.\cite{auretal1}.
A detailed study
was then made of the difference between the analytic and numerical Monte Carlo
{\it subtraction} pieces for various values of the isolation parameters
$\epsilon$ and $\delta$ at $\sqrt{S}=1$ TeV for the direct contribution. 
It was found that the
small cone approximation was within $10\%$ of the Monte Carlo results
for the subtraction piece except for very large values of $\epsilon$ and
$\delta$, greater than $0.25$ and $0.8$ respectively. This translated
into a {\it very small} error for the {\it full} isolated cross section even for
large values of the parameters, as the subtraction piece is numerically
much smaller than the inclusive piece. For example, at $p_T^{\gamma}=15$
GeV the subtraction piece is less than $10\%$ of the direct contribution
to the isolated cross section.     

\subsubsection{The Fragmentation Contribution}

A similar method was employed for the fragmentation contribution in NLO
and although a comparison with a Monte Carlo calculated version could not
be made, it was quite reasonably assumed that the result from the direct
study could be carried over, at least as far as numerical accuracy is
concerned. In this case the analytic method consists of imposing a cut
on the remnants of the fragmenting parton, just as in the LO case, by
imposing the restriction $z\geq {\rm MAX}[z_{min},1/(1+\epsilon)]$ on the
integration over $z$. In NLO this cut is no longer sufficient to isolate
the photon, since there is now a third parton in the final state which
may also be radiated into the isolation cone. In ref.\cite{gorvogel3}
this is dealt with by calculating a subtraction piece, similar to the
direct case which removes the contribution to the cross section whenever 
this third parton is in the cone, and is radiated collinearly to the
fragmenting parton, providing that the sum of its energy plus the energy of
the remnants of the fragmenting parton is above the
threshold, $\epsilon E_\gamma$. Thus the fragmentation contribution to
the isolated cross section is 
\begin{equation}
E_\gamma\frac{d\sigma^{isol}_{frag}}{d^3p_\gamma}=
E_\gamma\frac{d\sigma^{z\geq \frac{1}{1+\epsilon}}_{frag}}{d^3p_\gamma}-
E_\gamma\frac{d\sigma^{sub}_{frag}}{d^3p_\gamma},
\end{equation}
where the first term on the RHS is the NLO cross section with the
insufficient $z$-cut implemented and the second term is the cross
section with the additional parton in the cone. Schematically the
subtraction piece is given approximately by
\begin{equation}
E_\gamma\frac{d^3\sigma^{sub}_{frag}}{d^3p_\gamma}= \epsilon^a\left[
\ln\delta(A +A'\ln\delta) +B
+C\delta^2\ln\epsilon\right],
\end{equation}
 with new coefficients $A,A',B$ and $C$, all of which were calculated in
\cite{gorvogel3}.

In \cite{gorvogel3} it was not practical to test the accuracy of the
analytic method for isolating fragmentation contribution because of the
complexity of the NLO matrix elements. Its accuracy was therefore
assumed to be similar to the direct piece. This assumption is resonable,
but it would nevertheless be desirable to test it. Work is currently
underway to test the accuracy of the method by considering the less complex
photoproduction processes $\gamma q\rightarrow q g g$ and $\gamma
g\rightarrow q\bar{q} g$. The final states of these processes are the
same as those of $g q\rightarrow q g g$ and $g
g\rightarrow q\bar{q} g$ which are two of the processes found for the
hadron-hadron case. Any conclusion drawn about the accuracy of
the isolation method from these processes can be generalized to
the hadron-hadron case. For the purposes of the present work however, it
is assumed that the method is as accurate for fragmentation as it is for
the direct piece.

Before leaving this section it should be noted that according to
ref.\cite{bergerqui1} there is a serious problem with the definition of the 
isolated cross section in the fragmentation case. They concluded that 
conventional factorization breaks down in the fragmentation case when
isolation restrictions are imposed on the photon. Thus the cross section
can no longer be written in the schematic factorized form
\begin{equation}
\sigma=f_a(x_a,M^2)\ast f_b(x_b,M^2)\ast\hat{\sigma}^{ab\rightarrow
cX}\ast D_c^\gamma(z,M_f^2),
\end{equation}
where the hard scattering cross section, $\hat{\sigma}^{ab\rightarrow c
X}$ is infrared finite. They calculated the cross section for the process 
$e^+e^-\rightarrow \gamma+X$ and found it not to be infrared 
finite when the photon is isolated.

Arguments were also presented in ref.\cite{bergerqui1} for
the hadron-hadron case with the conclusion that it has an integrable
logarithmic divergence which renders the calculation unreliable. 
The cross section differentiable in a specific variable for the
subprocess $q\bar{q}\rightarrow q'\bar{q}'g$ with subsequent
fragmentation of one of the final state quarks was studied and found to
be larger than the corresponding inclusive cross section in the region
of phase space where a soft gluon is radiated into the isolation cone. 
This situation would be unphysical since the phase space for the isolated 
cross section is more restricted than for the non-isolated case. These 
conclusions were 
disputed in ref.\cite{auretal} where arguments were given to support
the view that the physical cross section is singularity free and thus
well defined. While it is not the purpose of this paper to support either
position for the general case, I shall nevertheless show that the 
calculation presented here for the fragmentation contribution in NLO
as calculated in \cite{gorvogel3} is well defined and regular in all
regions of phase space. In fact the large logarithms mentioned 
in ref.\cite{bergerqui1} are well behaved and regulated in the limit where 
the gluon becomes soft as they should be for any well defined cross section, 
and therefore cannot render the isolated 
cross section unreliable.   

My argument will rest on the following three points;
\begin{enumerate}
\item Integration over the kinematic variables $w$ and $z$ is necessary
in order to define the physical cross section whether the photon is
isolated or not. 
\item Only the subtraction piece of the cross section defined in
equation (2.10), namely $E_{\gamma}d\sigma^{frag}_{sub}/d^3p_{\gamma}$, can
potentially have any soft divergence associated with the isolation
procedure.
\item All potential soft divergences are well regulated in the
physically measurable cross section.
\end{enumerate}

Let us address point (1) by first noting that in this
calculation the soft gluon limit is the limit $w\rightarrow
1$ for {\it both} terms on the RHS of eq.(2.10), thus the soft gluon
limit is a clearly identifiable and unique point in the calculation as
it should be since the two parts are not independent but together make
up the isolated cross section. One can thus compare how the soft limit
is approached for each part and make sure that this point is neither over
counted or under counted in the definition of the isolated cross section. If the
point is counted twice then clearly extra soft singularities would be
created and the cross section would not be finite. If the point is not
counted at all then this would lead to uncanceled soft singularities in
the virtual piece with the same result. 

In the case of the first term of the right of eq.(2.10)
the soft divergences are regulated by `plus'-distributions such as $1/(1-w)_+$
which are defined to be regular in the soft limit for integration of a
test function $f(w)$, not singular at $w=1$, over $w$ by the equation
\begin{equation}
\int^1_A \frac{f(w)}{(1-w)_+} dw =\int^1_A\frac{f(w)-f(1)}{1-w}dw +
\ln(1-A)f(1). 
\end{equation} 
Thus the first term of eq.(2.10) is only defined and regular if an integration 
over $w$ is
performed over a finite region when $w$ is near $1$. Note also that although 
the divergences are regulated, near this point the region of integration 
cannot be taken too narrow because even without isolation the cross section 
would become infrared sensitive, i.e. if $A$ is too near to $1$ then
$\ln (1-A)$ would diverge. This type of infrared sensitivity is well 
known in perturbative QCD and can only be cured by soft gluon resummation. 
The lower limit of integration over $w$, $A$, is set by external
kinematics in the hadronic system and is not affected by the isolation
cuts. Clearly then the isolated cross section of eq.(2.10) is only defined 
when integration over  $w$ is performed. 

Point (2) is easily shown to be true. The first term on the right of
eq.(2.10), $E_{\gamma}d\sigma_{frag}^{z\geq\frac{1}{1+\epsilon}}/d^3p_{\gamma}$,
involves a cut on the variable $z$ only and this occurs after the soft
limit $w\rightarrow 1$ is regulated by the procedure defined by eq.(2.13). In 
other words, placing a cut on the remnants of the fragmenting quark alone does
not place any restrictions on the phase space available to the gluon.
This means no divergent logarithms associated with the isolation
procedure can come from this term. 

Point (3) is the most important and must be examined carefully.
To see where a divergence may potentially occur in our treatment of the
isolated cross section we must examine the structure of the subtraction
piece. If we assume that
a fragmenting quark with original energy $E_q$ and final energy $E_q'$
plus a gluon with energy $E_g$ are inside the cone, then the isolation
condition is 
\begin{equation}
E_g+E_q'> \epsilon E_\gamma
\end{equation}
for the subtraction piece. 
This translates to a condition 
\begin{equation}
z(1+\epsilon)< \frac{1}{1-v+v w}
\end{equation} 
on the phase space for the subtraction piece. In ref.\cite{gorvogel3},
in the small cone limit, the subtraction cross section was written as
\begin{eqnarray}
E_\gamma\frac{d\sigma^{sub}_{frag}}{d^3p_\gamma}&=&\frac{1}{\pi
p_T^2S}\sum_{abc}\int^1_{max(1-V+V
W,\frac{1}{1+\epsilon})}dz\int^{1-(1-V)/z}_{V W/z}dv\int^1_{V W/z v}dw
\nonumber \\
&\times &
f^a_1(x_1,M^2)f^b_2(x_2,M^2)D^\gamma_c(z,M_f^2)\hat{s}\frac{d\hat{\sigma}_
\delta^{ab\rightarrow c}}{dv dw}\Theta\left( \frac{1}{1-v+v
w}-z(1+\epsilon)\right),
\end{eqnarray}
where the kinematic variables are defined as before, and
$d\hat{\sigma}^{ab\rightarrow c}_\delta/dv dw$ is the hard cross section
in the small cone approximation. Consider, for example, the process
$qg\rightarrow q X\rightarrow\gamma X'$ where the photon is
produced by fragmentation off a quark. In the small 
cone/collinear limit the hard cross section is given by
\begin{eqnarray}
\frac{d\hat{\sigma}_\delta^{qg\rightarrow q}}{dv
dw}&=&\frac{\alpha_s}{2\pi}\frac{v}{1-v+v w}\nonumber \\
&\times &\left[P_{qq}(1-v+v
w)\frac{d\hat{\sigma}^{qg\rightarrow qg}(\hat{s},y)}{dy}+P_{qg}(1-v+v
w)\frac{d\hat{\sigma}^{qg\rightarrow gq}(\hat{s},y)}{dy}\right],
\end{eqnarray}
where $y=v w/(1-v+v w)$, and the splitting functions are given by
\begin{eqnarray}
P_{qq}(z)&=&C_F\left[ \frac{1+z^2}{1-z}\ln\left(
\frac{v^2(1-w)^2E_\gamma^2\delta^2}{z^2 M_f^2}\right)+(1-z)\right]\nonumber \\
P_{qg}(z)&=&\left[\frac{z^2+(1-z)^2}{z}\ln\left(
\frac{v^2(1-w)^2E_\gamma^2\delta^2}{z^2 M_f^2}\right)+z(1-z)\right].
\end{eqnarray}
The splitting function $P_{qq}(z)$ where $z=1-v+v w$ has a $1/(1-w)$
behavior which, after integration over $w$ behaves as $\ln(1-w)$. In
eq.(2.16), the upper limit of the $w$ integration is 
\begin{displaymath}
1-\frac{1}{v}+\frac{1}{zv(1+\epsilon)},
\end{displaymath}
which is $1$ if $z=1/(1+\epsilon)$. This occurs when the energy of the
gluon inside the cone, $E_g=0$ and therefore $E_q'=\epsilon E_\gamma$,
i.e., at the lower limit of the $z$ integration in eq.(2.16), the logarithm 
$\ln[(1-(1+\epsilon)z)/((1+\epsilon)v z)]$ becomes $\ln[0]$.
This apparent logarithmic divergence, if not regulated, could render the 
subtraction piece, and therefore the full fragmentation contribution to the 
isolated cross section unreliable. We must therefore examine how the
subtraction piece actually behaves in this limit. 

From a physical point of view we do not want to subtract the
contribution to the cross section when $E_g=0$ inside the cone as this passes 
the isolation cut and hence is part of the measured cross section. This point 
is included in the first term on the RHS of
eq.(2.10) where the poles have been cancelled in $4-2 \epsilon$
dimensions and all attendant divergences have been regulated by the
`plus'-distributions. Thus when this limit is approached in the 
subtraction piece, the subtraction piece should give no 
contribution and should ideally vanish. In an approximate 
calculation the subtraction piece may still give a (small) finite contribution
in this limit, but there should be no divergent logarithms present.  
This is the justification for doing the rest of the phase space integrals in 
$4$-dimensions. We thus need to ensure that there are no divergent
logarithms in the soft limit of the subtraction piece that would render the
predictions unreliable. In fact the potential singularity is actually 
regulated and is therefore {\it not} dangerous as long as we integrate over 
the variables $w$ and $z$, as required for the definition of the
physical cross section.

From eqs.(2.17) and (2.18) the term proportional to $1/(1-w)$ would be
the first potentially dangerous term. Let us for the moment ignore the
dependence of the structure and fragmentation functions on the variables
$z$ and $w$. In fact the fragmentation function is actually small at the
point $z=1/(1+\epsilon)$ which is near to $z=1$ where it vanishes. 
It turns out that when we integrate over $w$ and $z$ this singular term becomes 
\begin{equation}
\left[
\left(\frac{1}{1+\epsilon}-z\right)\ln[z(1+\epsilon)-1]+z\ln[z(1+\epsilon)]
\right]^1_{\frac{1}{1+\epsilon}}
\end{equation}
which clearly vanishes in the limit $z\rightarrow 1/(1+\epsilon)$ as it
should. A similar situation occurs for the term $\ln(1-w)/(1-w)$. The conclusion
is thus that there are no unregulated soft divergences in the cross section 
defined by eq.(2.16), and thus also not in that of eq.(2.10). 
This behaviour is supported by the results of the numerical calculation
where the integral converges smoothly to a finite limit.

One may argue that if the integrand as a function of $w$ is singular when 
$w\rightarrow 1$ then the predictions could be unreliable. If one were
to plot a curve of $d{\hat{\sigma}}/{dvdw}$, the hard subprocess cross section,
versus $w$, whenever $w=1$ a singularity would be encountered. Of
course this singular point would also be encountered for the fully inclusive 
subprocess cross section. As stated above the hard subprocess cross section in 
both the 
inclusive and isolated cases are defined in terms of `plus'-distributions such 
as $1/(1-w)_+$ and $[\ln(1-w)/(1-w) ]_+$ which have no meaning at $w=1$
unless one integrates over a finite region including this point to
define a physically observable cross section. Thus
plotting the hard subprocess cross sections vs $w$ is meaningless as no
conclusions could be drawn from the result.
 
The conclusion is therefore that although it is generally accepted that 
further work is needed in order to fully deal 
with isolation in the fragmentation case at both leading and next to
leading order, mostly to do with defining the fragmentation functions
when the isolation restrictions are imposed, for the well defined
physical cross section presented here there are no soft divergences
which survive to make the predictions unreliable.

\subsubsection{ The Monte Carlo Method}

The Monte Carlo method was first used in \cite{baeretal} for the unpolarized 
case and a detailed description can be found there so only an outline is
presented here. Some details are also presented in ref.\cite{gordon}
where the polarized case is highlighted. When the two-to-three hard
scattering subprocesses are integrated over phase space, singularities
are encountered whenever a gluon becomes soft or two partons become
collinear. Some of these singularities will cancel against similar ones
in the virtual two-body contributions. The rest are factored into the
parton distribution and fragmentation functions at some arbitrary scale,
or cancelled when an appropriate jet definition is implemented. 
The Monte Carlo technique consists essentially in identifying those
regions of phase space where soft and collinear singularities occur and
integrating over them analytically in $4-2\varepsilon$ dimensions,
thereby exposing the singularities as poles in $\varepsilon$. 
In order to isolate these regions from the rest of the three-body phase space 
arbitrary boundaries are imposed through the introduction of cutoff
parameters, $\delta_s$ and $\delta_c$. The soft gluon region of phase space is 
defined to be the
region in which the gluon energy, in a specified reference frame, usually
the subprocess rest frame, falls below a certain threshold,
$\delta_s\sqrt{\hat{s}}/2$, where $\delta_s$ is the cutoff
parameter, and $\hat{s}$ is the center-of-mass energy in the initial
parton-parton system. Labelling the momenta for the general three-body
process by $p_1+p_2\rightarrow p_3+p_4+p_5$, the general invariants are defined
by $s_{ij}=(p_i+p_j)^2$ and $t_{ij}=(p_i-p_j)^2$. The
collinear region is defined as the region in which the value of an invariant
falls below the value $\delta_c \hat{s}$. 

Specific approximate versions of the full three-body matrix elements are
integrated over the soft and collinear regions of phase space.
For the soft-gluon case the approximate matrix element 
is obtained by setting the gluon energy to zero everywhere 
it occurs in the matrix elements, except in the denominators. For the 
collinear singularities, each 
invariant that vanishes is in turn set to zero everywhere except in the
denominator. This form is the leading pole approximation. These approximate 
expressions are then integrated over phase space and only the terms 
proportional to logarithms of the cutoff parameters are retained. All terms 
proportional to positive powers of the cutoff
parameters are set to zero. Therefore in order for the method to yield reliable 
results, the cutoff parameters must be kept small, otherwise the 
approximations would not be valid.  

Once the two-to-three particle phase space integrals are performed 
analytically over the singular regions exposing the soft and collinear poles, 
the $O(\alpha ^2_s)$ virtual gluon-exchange loop contributions, if these
are present, are added, and all double and single poles are verified to cancel, 
as they should. The remaining collinear 
singularities are factored into parton distribution and fragmentation 
functions at an appropriate factorization or fragmentation scale, after
an appropriate factorization scheme is chosen. In this calculation I use
the $\overline{MS}$ scheme. One is now left with a set of matrix 
elements for effective two-body final-state processes that depend explicitly 
on $\ln\delta_s$ and $\ln \delta_c$.  In addition, the non-singular regions of 
phase space yield a set of three-body final-state matrix elements
which, when integrated over phase space by Monte Carlo techniques,
have an implicit dependence on these same logarithms. The signs are such 
that the dependences on $\ln\delta_s$ and $\ln \delta_c$ cancel between 
the two-body and three-body contributions. The physical cross sections are
independent of these arbitrary cutoff parameters over wide ranges.  
All the two and three-body matrix elements for both the
polarized and unpolarized calculation are presented in the appendix of
ref.\cite{gordon}.

The Monte Carlo method outlined above is well suited to calculations of
processes involving well defined experimental selections such as isolation 
cuts.  Since all singularities have already been cancelled, whenever an event 
fails an experimental cut it is simply rejected and does not contribute to the
cross section. As emphasized earlier, at colliders it is necessary to 
restrict the hadronic
energy allowed near the photon in order that the signal may be
identified unambiguously. In the method of calculation just outlined
here the isolation cuts can be implemented exactly.

\section{Numerical Results}

In this section numerical results are presented for the 
cross section at a center-of-mass energy of $\sqrt{s}=1.8$ TeV
appropriate for the CDF and D0 experiments at Fermilab.
First a comparison is made between the cross sections
calculated using purely analytic methods and using the Monte Carlo
technique. In these comparisons the fragmentation contributions are
included with LO matrix elements only, whereas the direct ones are always 
calculated in NLO. Later when a comparison with the CDF data is made then the 
fragmentation contribution calculated using the approximate analytic method at 
NLO is included. 
In all cases where LO matrix elements are used, the parton
densities and fragmentation functions evolved in NLO are nevertheless used.

The renormalization, factorization, and fragmentation scales are always set to 
a common value $\mu = p_T^{\gamma}$ unless otherwise stated. The
fragmentation functions evolved in NLO from ref.\cite{vogtfrag} are used
throughout, but various parametrizations of the parton densities of the
proton are used. The NLO two-loop expression for $\alpha_s$ is always used and
four quark flavors are assumed $(N_f=4)$. The value of $\Lambda$ used is 
chosen to correspond with the parton densities in use.   

\subsection{Analytic vs Monte Carlo Isolation Methods}

The Monte Carlo and analytic methods of calculating the inclusive cross
section should give the same results for the same choice of parameters.
This can serve as a check on both calculations. In fig.1a the
predictions are compared for the standard choice of parton distributions
in this subsection, the CTEQ3M set \cite{CTEQ3} at $\sqrt{s}=1.8$ TeV. Clearly 
they agree over many decades of $p_T^{\gamma}$ as they should. In fact
the numerical differences are generally less than $4\%$ and are always
within the errors of the Monte Carlo calculation. For the
isolated cross section, the analytic calculation is only strictly valid
for a very small cone and small values of the energy resolution
parameter $\epsilon$, so it is interesting to see whether it gives similar
predictions to the Monte Carlo for CDF parameters, $R\sim \delta=0.7$ and
$\epsilon=2{\rm GeV}/p_T^{\gamma}$. Fig.1b compares the isolated cross section
as predicted by the analytic and Monte Carlo calculations. Form this
graph it can be seen they agree extremely well at all $p_T^{\gamma}$ values
shown. Again at no point is there a difference of more than $\pm 3-4\%$
between the two, and the differences are always within the Monte Carlo errors.
Fig.1c shows the predictions of the Monte Carlo calculation
broken down into some of the subprocess contributions. This graph
confirms that the cross section is dominated by initial $qg$ scattering
but also shows that $q\bar{q}$ scattering is important especially in the
large $p_T^{\gamma}$ region.

In figs.2a and 2b the dependences on $\epsilon$ and $R$ are compared
at two values of $p_T^{\gamma}$.
In the case of $\epsilon$ dependence the results are very similar
except at very small values of $\epsilon$ where the analytic results
tend to be larger. This feature was noted in ref.\cite{gorvogel3} where
it was shown that the analytic calculation is more accurate at lower
$p_T^{\gamma}$. For the $R$ dependence there is a larger discrepancy
between the two curves for $R\leq 0.3$. It seems that the analytic
calculation gives results which are too large in this region, and for
$R\leq 0.1$ gives results which are even larger than the inclusive one.
Clearly the analytic approximation is breaking down at this point. This
is understandable since only terms such as $\ln\delta$ are retained which
give a large negative contribution to the subtraction piece of the cross
section. Thus there appears to be a point below which the analytic
approximation does not give reliable results, despite the fact that in
principle it should be valid for a small cone. For all values of $R$
above $0.4$ the two methods give essentially identical results, hence
the small cone approximation should be reliable for CDF isolation 
parameters.   

\subsection{Comparison with CDF Data}

Now that we have established that the analytic and Monte Carlo methods
give essentially the same results for the isolated cross section at CDF
we are in a position to compare predictions with data. An interesting question
concerning the NLO calculation as presented here is how much of the
cross section is due to the fragmentation contribution, and how much
enhancement does the inclusion of the NLO fragmentation contributions
bring. To answer this question, in fig.3a the ratio
\begin{displaymath}
\frac{\sigma^{frag}}{\sigma^{dir}+\sigma^{frag}}
\end{displaymath}
is plotted for the isolated cross section with CDF parameters.
$\sigma^{dir}$ is the direct contribution to the cross section in NLO,
whereas $\sigma^{frag}$ is calculated both with LO (dashed line) and
with NLO matrix elements (solid line). As one may expect, the
fragmentation contributions are most important at low $p_T^{\gamma}$.
When it is calculated in NLO the fragmentation contribution clearly
gives a larger contribution to the cross section, indicating that a
better agreement with the CDF data, which are generally steeper than the
theoretical predictions may be expected in this case. To further
highlight this point, in fig.3b the ratio
\begin{displaymath}
\frac{ \sigma^{dir}+\sigma^{frag}(NLO) }{
\sigma^{dir}+\sigma^{frag}(LO)}
\end{displaymath}
is plotted for both the isolated and non-isolated cases. The denominator
corresponds to predictions made by the calculation of
ref.\cite{baeretal}, where the fragmentation contributions are included
with LO matrix elements. The NLO fragmentation contributions give up to
$15\%$ enhancement to the cross section at low $p_T^{\gamma}$ for the
isolated case.
Comparison with the non-isolated case indicates that the effect of the
NLO fragmentation contributions are suppressed relative to the LO ones
after isolation cuts are imposed. This is not surprising since isolation
involves a more severe restriction on the three-body phase space of the
NLO contributions but similar restrictions on the two-body phase space
parts of both the LO and NLO  calculations.

In fig.4a a comparison is made with the CDF data for the isolated cross
section and the full NLO calculation using various parton densities
labelled MRSR \cite{MRS96}, CTEQ3M \cite{CTEQ3} and GRV94 \cite{GRV94}. 
In the case of the GRV94 distributions, no charm contribution was
included in the parametrization, the intention being that the contribution due 
to charm quark
scattering be calculated using matrix elements that explicitly take the
charm quark mass into account. Since these matrix elements are not yet
available in NLO, the GRV94 curve has been supplemented by including a
charm density from their older parametrization, \cite{GRV91}. This
enhances the cross section by about $20\%$ in the low $p_T^\gamma$
region and allows an easier shape comparison with the curves from the
other parton densities. From fig.4a it appears that all the curves agree
reasonably well with the data, with MRSR4 perhaps giving a better
description of the data in the low $p_T^\gamma$ region. Fig.4b which
highlights this problematic region confirms this.

A more detailed comparison with the data is made in fig.4c where the
quantity
\begin{displaymath}
\frac{DATA-THEORY}{THEORY}
\end{displaymath}
is plotted. If data and theory agreed exactly then this line would
ideally be a straight line through $0$. The data is compared directly to
the MRSR2 parametrization, and taking this set as the default, the
results using the other sets are substituted for the data and also
plotted in the figure. A similar procedure is carried out in fig.4d, but there 
taking the CTEQ4HJ set, which is fitted
to the new CDF large $p_T$ jet data \cite{CTEQ4} as the default. There is a
substantial spread in the curves and taken as a whole the band they
provide describes the data well. But taken individually they clearly
all leave some slope difference with the data taken as a whole. This slope 
is substantially diminished when
compared to most previous comparisons with the data \cite{abe},
partly due to the enhancement from the fragmentation and partly due to
the modern, steeper parton distributions. Even recently there was up to
$40\%$ difference between the theoretical curves and the data at low
$p_T^\gamma$ whereas now, for the steepest distributions, this is
reduced to under $10\%$. The overall slope in the data taken over all
$p_T^\gamma$ values for the MRSR2 parametrization is roughly $20\%$,
whereas the systematic uncertainty as reported by CDF is of order
$5\%$.

A change in the factorization/renormalization scales from
$\mu^2=(p_T^\gamma)^2$ to $\mu^2=(p_T^\gamma/2)^2$ does not affect the
slope of the curves substantially in the $p_T$ range considered here.
The major effect of this change in the scale is a normalization shift of
all the curves upwards. This uncertainty in the theoretical predictions
therefore cannot be exploited to explain the slight slope difference
between the data and theory. Although the parton distributions are much
better known than they were a few years ago, as the differences between the
various curves clearly show, there is still enough uncertainty here, 
particularly in the
gluon distributions, to explain at least some of the slope differences.
Another area which has also been explored with some success
\cite{kuhlmann} is inclusion of  
initial state transverse momentum $k_T$ effects due to soft gluon radiation 
off the initial state partons. There is also the possibility
that estimates of the NLO fragmentation contributions could be improved
via either better theoretical treatment of the isolation conditions
on the matrix elements or on the fragmentation functions themselves. But
although this may have some effect on the predictions at low
$p_T^{\gamma}$ it will not affect substantially the predictions above 
$p_T^{\gamma}=30-40$ GeV, where the theory over predicts the data by an
almost constant $10\%$. 

An interesting and rather curious thing to
observe about the discrepancy between the theoretical predictions and
the data is that if one were to separate the high ($\geq
30$) GeV and the the low $p_T^{\gamma}$ regions then the theoretical
predictions could be made to fit the data in these regions separately
with different structure functions. For example MRSR4 describes the
low $p_T^{\gamma}$ data extremely well (fig.5a), whereas MRSR1 does very well
at high $p_T^{\gamma}$ (fig.5b). Of course if the data is examined between
$p_T^{\gamma}\sim 15$ and $30$ GeV (fig.5c) there appears to be an obvious slope
difference with the theoretical predictions that none of the  individual 
structure functions can fit. This could be taken as evidence that NLO QCD is
inadequate to describe the data in detail, and that new theory such as
the inclusion of soft gluon radiation is needed. 

\section{Conclusions}

The best state of the art calculation of isolated prompt photon
production at NLO was made in this paper. Fragmentation contributions were
included fully at NLO and the very latest parametrizations of the
proton distributions were used to make comparisons with the CDF data.  
These comparisons show that there is still some discrepancy between the
theoretical predictions and the data which must be addressed before the
cross section can be used to extract information on the gluon densities.
This discrepancy is now substantially reduced as compared to previous
studies, due to a combination of improvements in our knowledge of the parton
densities and inclusion of the fragmentation contributions at NLO. 

There is still sufficient spread in the predictions from different
parametrizations of the parton densities to indicate that further
improvements may be made in this area that would improve the comparison
between theory and data. Improvements in the theoretical predictions
could also be made by a better understanding of the fragmentation
contributions in NLO. In addition, inclusion of soft gluon radiation
could also be a possible way of improving the theoretical predictions.

A detailed comparison of two methods of calculating the inclusive and
isolated prompt photon cross sections, one using purely analytic
techniques and the other using the analytic/Monte Carlo technique was made 
confirming that the two methods give the same results. In the process a
previously unexplained discrepancy between the results of the calculations of 
refs.\cite{baeretal} and \cite{reya} was resolved.

\section{Acknowledgments}

I would especially like to thank S. Kuhlmann for running the code of
ref.\cite{baeretal}, and E. L. Berger and E. Kovacs for helpful discussions.  
I acknowledge Bob Bailey's help when I was originally learning to implement 
the Monte Carlo method used in this paper. Special thanks goes to thank Marie 
Maslowski for friendship and support while carrying out this work. This work 
was supported by the US Department of Energy, Division of High Energy Physics, 
Contract number W-31-109-ENG-38.
\pagebreak


\pagebreak
\noindent
\begin{center}
{\large FIGURE CAPTIONS}
\end{center}
\newcounter{num}
\begin{list}%
{[\arabic{num}]}{\usecounter{num}
    \setlength{\rightmargin}{\leftmargin}}

\item (a) Comparison of the inclusive prompt photon cross section
plotted vs $p_T^{\gamma}$ as predicted by the analytic \cite{gorvogel3} and 
analytic/Monte Carlo calculations at $\sqrt{s}=1.8$ TeV, and $\eta=0$. (b) 
Same as (a) but for the isolated cross section with $R=0.7$ and
$\epsilon=2\;{\rm GeV}/p^{\gamma}_T$. (c) The isolated cross section in
(b) broken down into contributions from initial state subprocesses.  
\item (a) Dependence of the isolated cross section on the energy
resolution parameter $\epsilon$ at fixed
$R=0.7$ for two values of photon $p_T$ as predicted by the analytic and
Monte Carlo calculations. (b) Dependence of the isolated cross section
on the isolation cone size $R$ for fixed $\epsilon$ for the analytic and
Monte Carlo calculations.
\item (a) Ratio of the fragmentation contribution to the isolated cross
section to the full isolated cross section, where the fragmentation 
contribution is calculated using LO and NLO matrix elements. 
(b) Ratios of the full NLO cross section to the NLO cross section
calculated with LO matrix elements for the fragmentation contribution
for the isolated and inclusive cases.
\item (a) Comparison of the full isolated cross section calculated using
different parametrizations of parton densities with the CDF data. (b)
Same as (a) but for the lower $p_T^{\gamma}$ region only. (c) and (d)
Plots of the ratio (DATA-THEORY)/THEORY, where THEORY is the full NLO
theoretical predictions calculated using the MRSR2 and CTEQ4HJ parametrizations
of the proton densities, and DATA are data from the CDF collaboration. The
different textured curves are obtained by substituting theoretical
predictions made using different parametrizations of the proton
densities for the DATA. 
\item (a) Plot of (DATA-THEORY)/THEORY where THEORY is the theoretical
predictions obtained using the MRSR4 proton densities in the low
$p_T^{\gamma}$ region of the CDF data. (b) Same as (a) in the higher 
$p_T^{\gamma}$ region for THEORY given by the MRSR1 proton densities. (c) as 
(a) and (b) for the central $p_T^{\gamma}$ region for THEORY given by
both the MRSR1 and MRSR4 proton densities.    
\end{list}
\pagebreak


\begin{references}
\bibitem{owens} J.F. Owens, Rev. Mod. Phys. {\bf 59}, 465 (1987).
\bibitem{bergerqui1} E. L. Berger, X. Guo and J.-W. Qiu, Phys. Rev.
Lett. {\bf 76}, 2234, (1996); Phys. Rev. {\bf D 54}, 5470, (1996).
\bibitem{auretal} P. Aurenche et. al., ENSLAPP-A-595/96, LPTHEORSAY
96-40, hep-ph/9606287.
\bibitem{abe} CDF Collaboration, F. Abe {\it et al.}, Phys. Rev. {\bf D 48},
2998 (1993); Phys. Rev. Lett. {\bf 68}, 2734 (1992). 
\bibitem{auretal1} P. Aurenche, R. Baier and M. Fontannaz, Phys. Rev.
{\bf D42}, 1440 (1990).
\bibitem{auretal2} P. Aurenche, R. Baier, A. Douiri, M. Fontannaz, and
D. Schiff, Phys. Lett. {\bf 140B}, 87 (1984); P. Aurenche, R. Baier, 
M. Fontannaz, and D. Schiff, Nucl. Phys. {\bf B297}, 661 (1988).
\bibitem{baeretal}  H.~Baer, J.~Ohnemus, and J.~F.~Owens, Phys. Rev. {\bf D42},
 61 (1990);B.~Bailey, J.~Ohnemus, and J.~F.~Owens, Phys. Rev. {\bf D46}, 
2018 (1992).
\bibitem{bergerqiu4} E. L. Berger and J. Qiu, Phys. Lett. {\bf B 248},
371 (1990); Phys. Rev. {\bf D 44}, 2002 (1991).
\bibitem{gorvogel3} L. E. Gordon and W. Vogelsang Phys. Rev. {\bf D50},
1901 (1993). 
\bibitem{reya} M.~Gl\"{u}ck, L.~E.~Gordon, E.~Reya, and W.~Vogelsang, Phys. 
Rev. Lett. {\bf 73}, 388 (1994).
\bibitem{gorvogel} L. E. Gordon and W. Vogelsang Phys. Rev. {\bf D48},
3136 (1993) and {\bf D49}, 170 (1994).
\bibitem{aversa} F. Aversa, P. Chiappetta, M. Greco and J.Ph. Guillet,
Phys. Lett. {\bf B 210}, 225 (1988); 211, 465 (1988); Nucl. Phys. {\bf
B327}, 105 (1989).
\bibitem{vogtfrag} M. Gl\"{u}ck, E. Reya and A. Vogt, Phys. Rev. {\bf D48},
116 (1993).
\bibitem{experiment} CDF Collaboration, F. Abe {\it et al.}, Phys. Rev. Lett.
{\bf 73}, 2662 (1994).
\bibitem{kuhlmann} J. Huston {\it et al.}, Phys. Rev. {\bf D 51}, 6139 (1995).
\bibitem{gordon} L. E. Gordon ANL-HEP-PR-96-59.
\bibitem{reno} H. Baer and M. H. Reno, Phys. Rev. {\bf D 54}, 2017
(1996).
\bibitem{CTEQ3} H. L. Lai {\it et al.,} CTEQ Collaboration, Phys. Rev.
{\bf D51}, 4763 (1995).
\bibitem{MRS96} A. D. Martin, R. G. Roberts and W. J. Stirling,
DPT/96/44.
\bibitem{GRV94}M. Gl\"{u}ck, E. Reya and A. Vogt, Z. Phys. {\bf C67},
433 (1995).
\bibitem{GRV91} M. Gl\"{u}ck, E. Reya and A. Vogt, Phys. Rev. {\bf D45},
3986 (1992).
\bibitem{CTEQ4} H. Lai {\it et al.}, CTEQ Collaboration,
MSUHEP-60426,CTEQ-604, hep-ph/9606399.
\end{references}
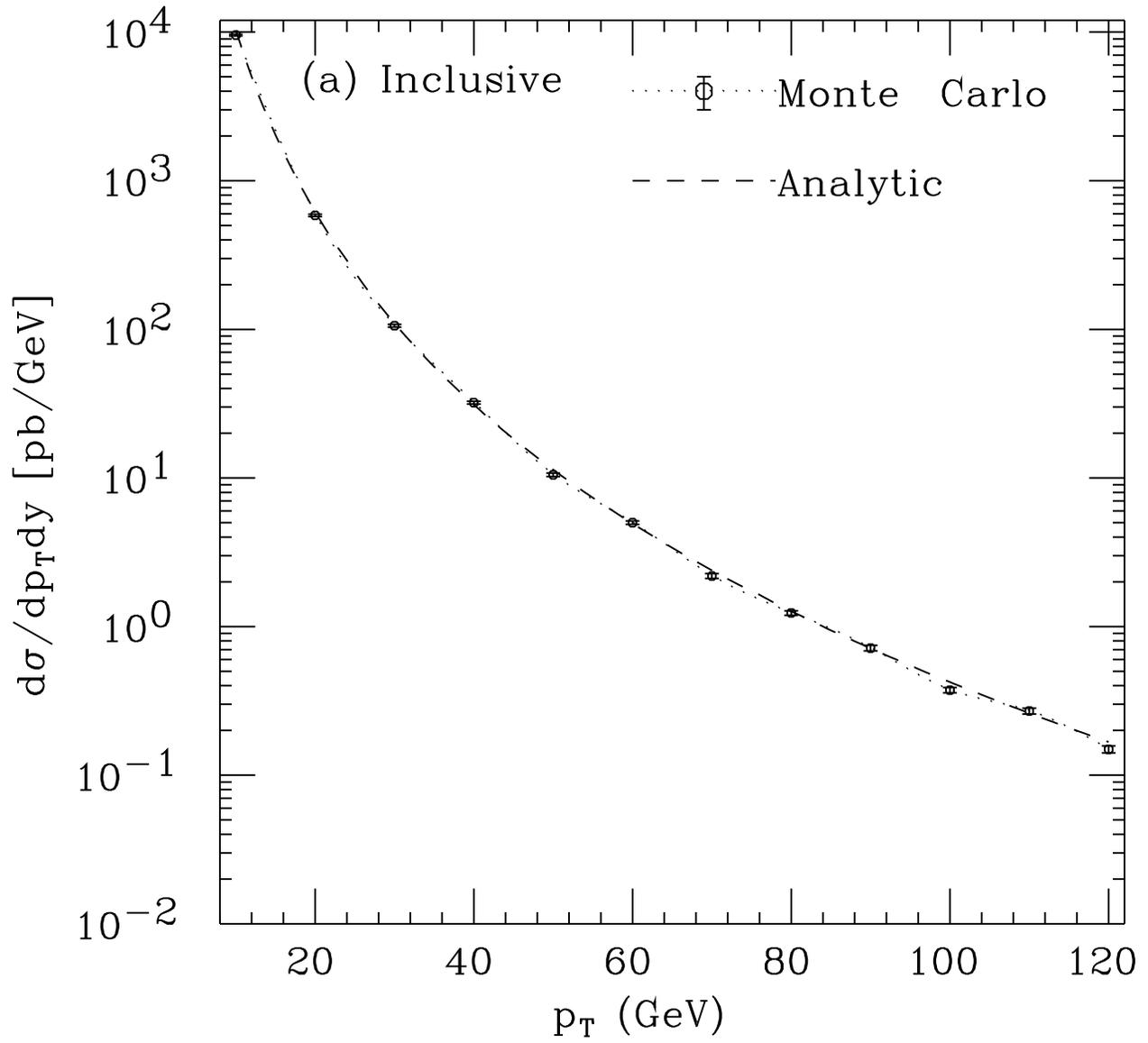
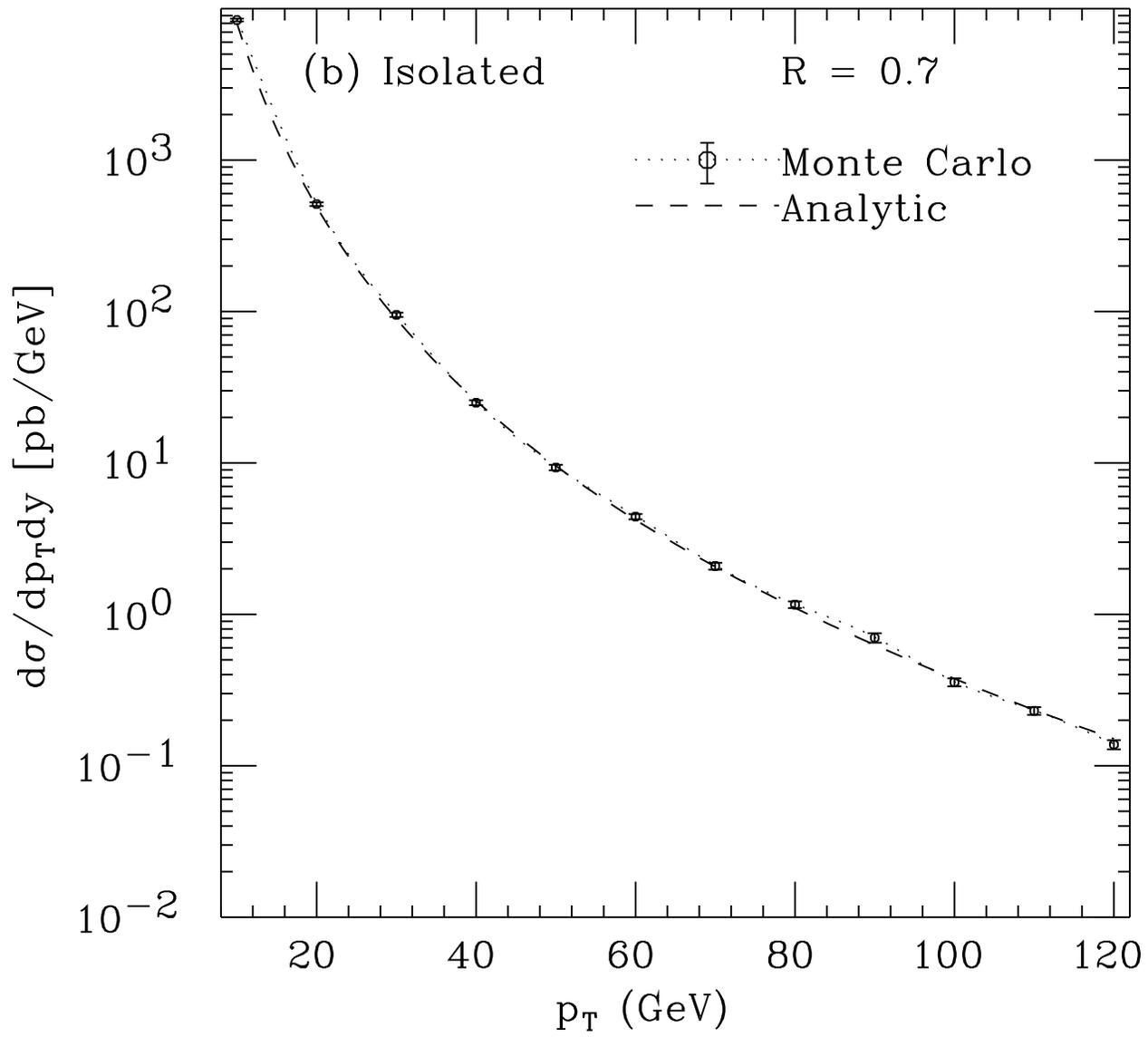
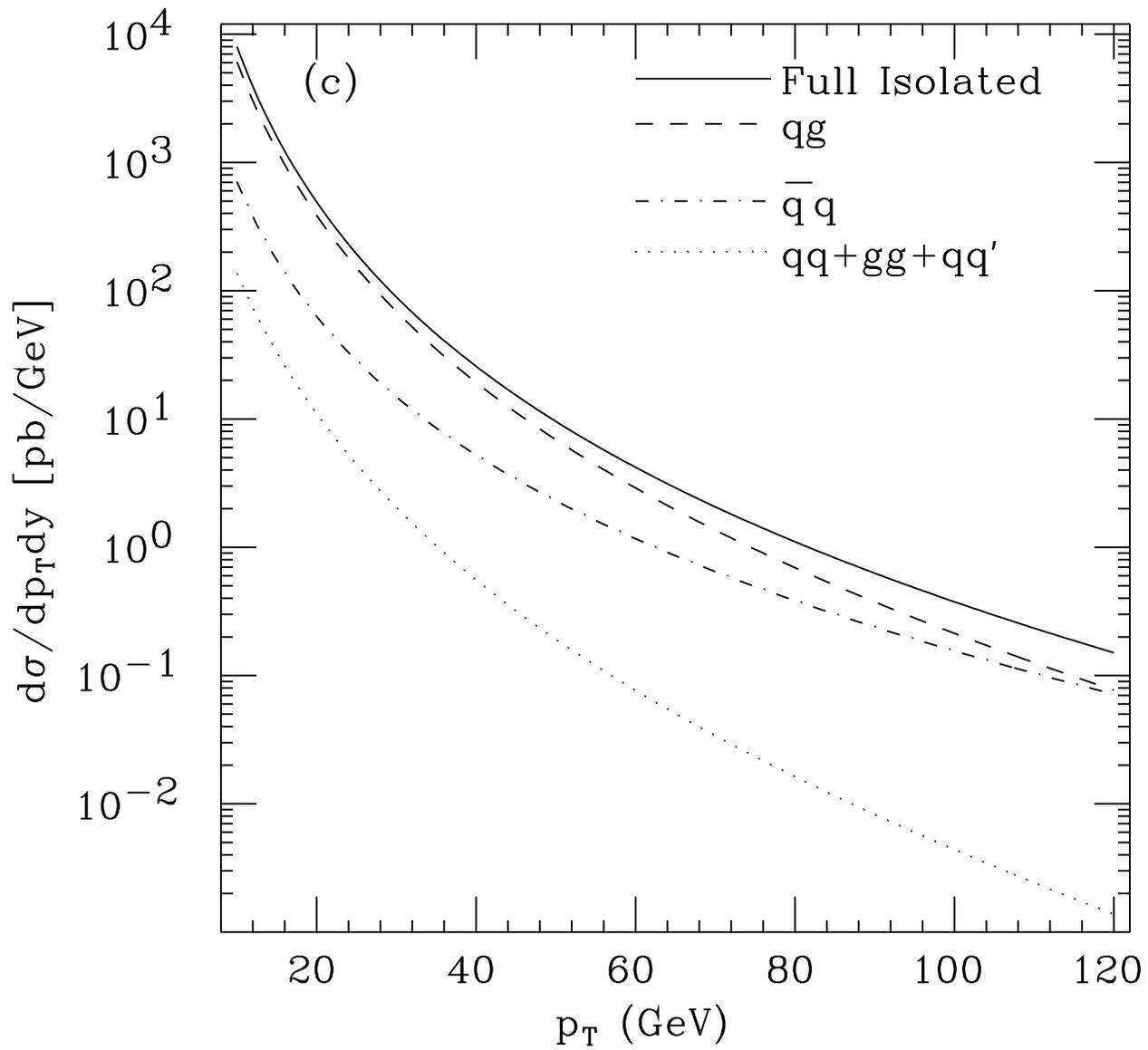
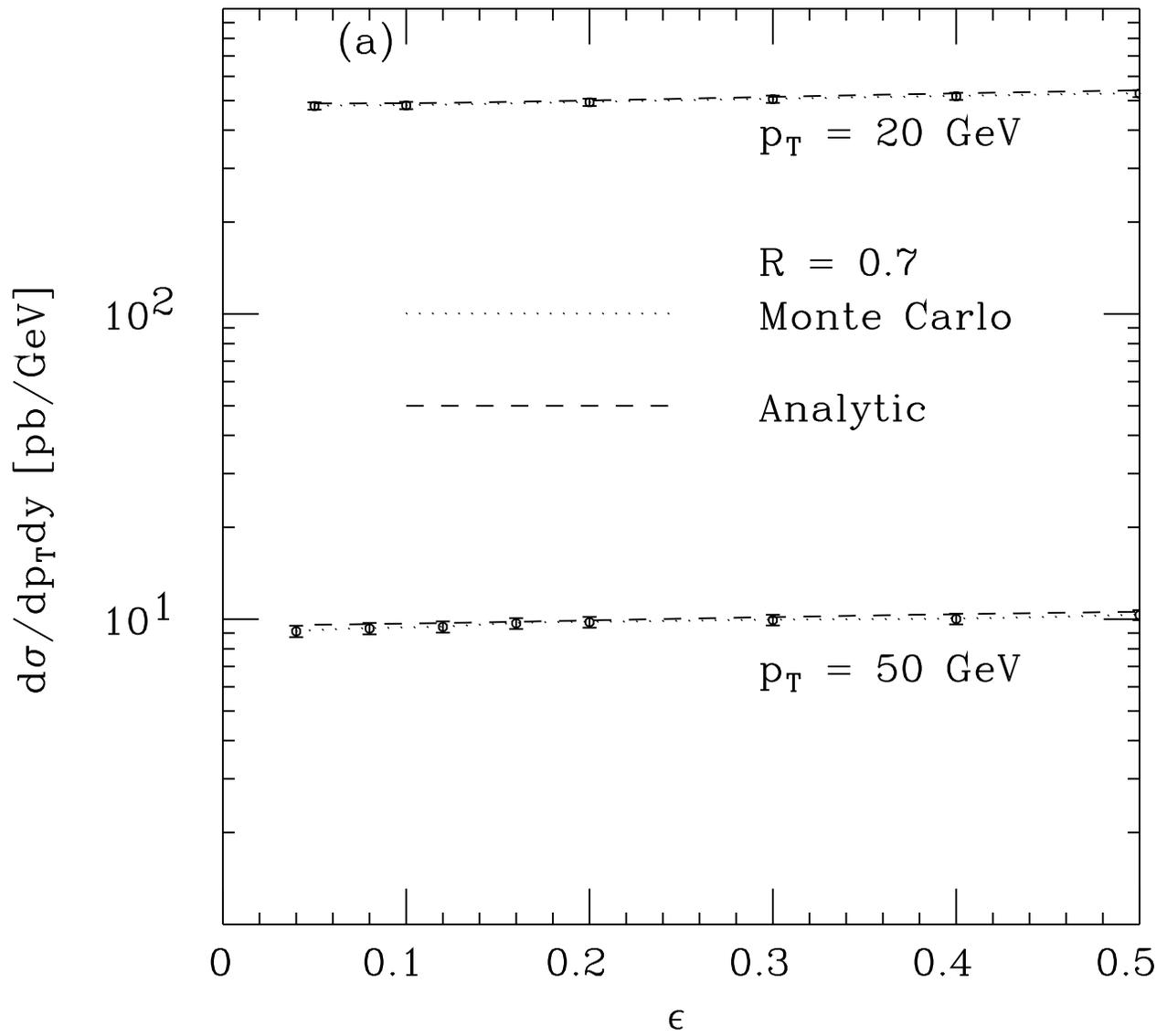
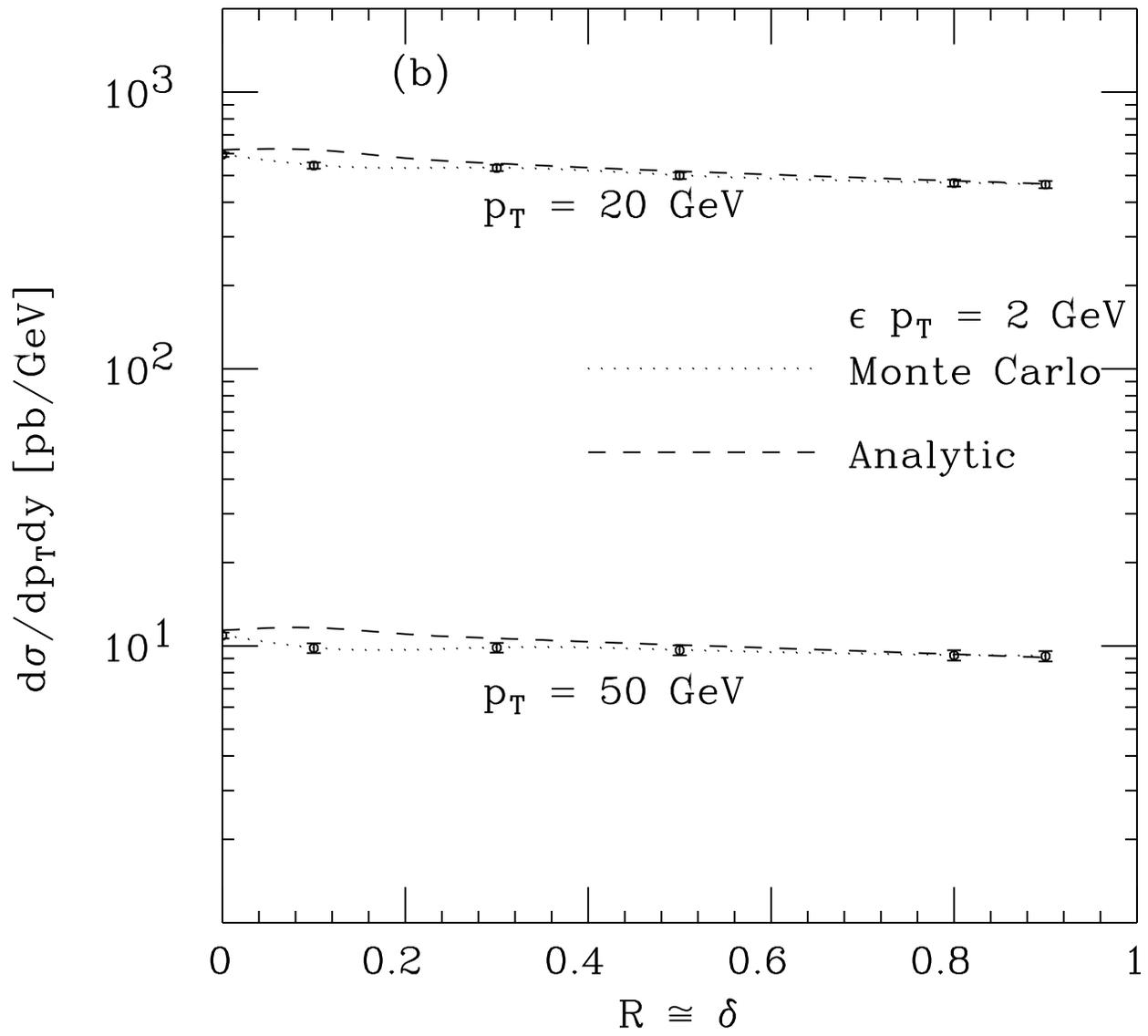
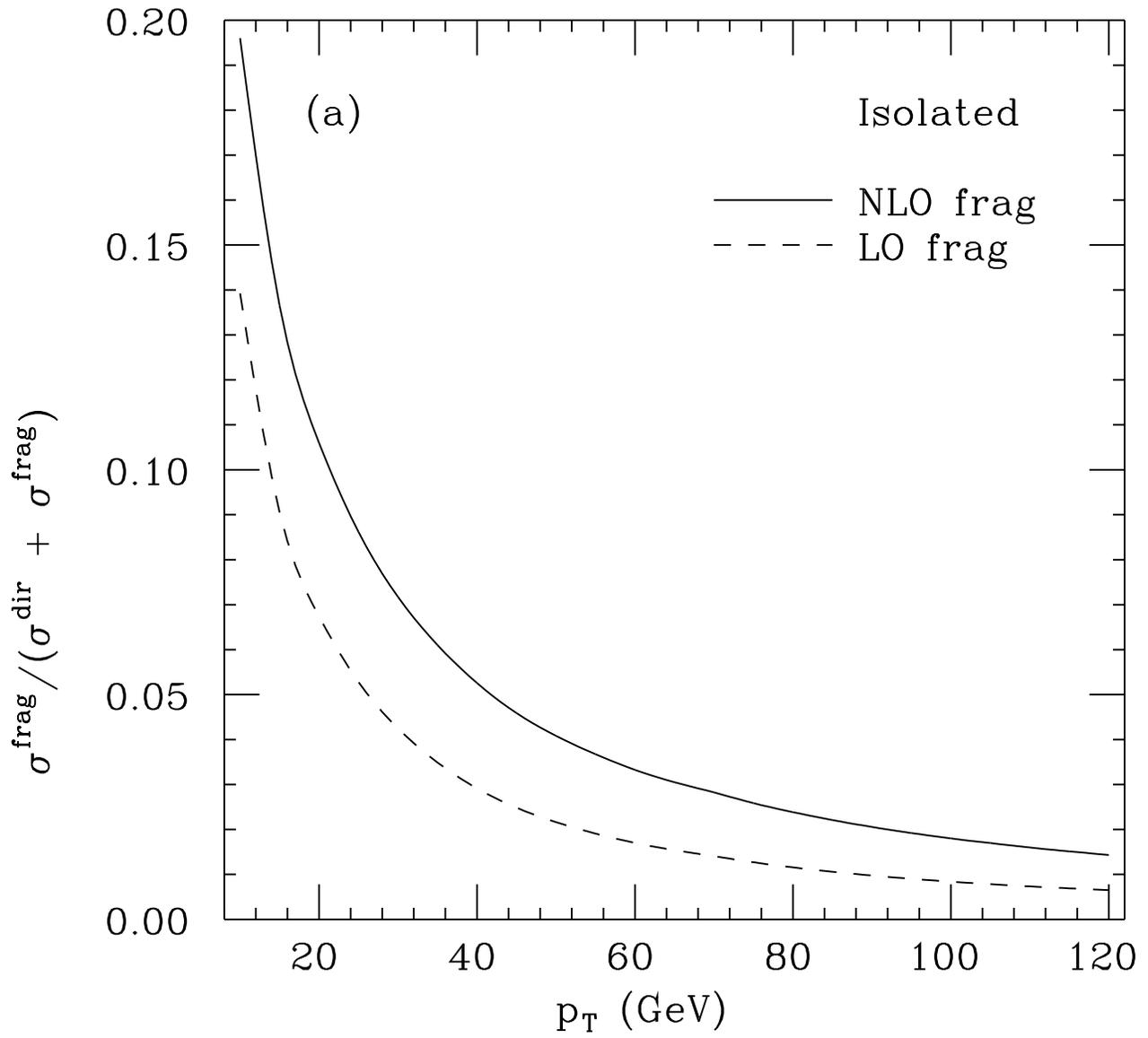
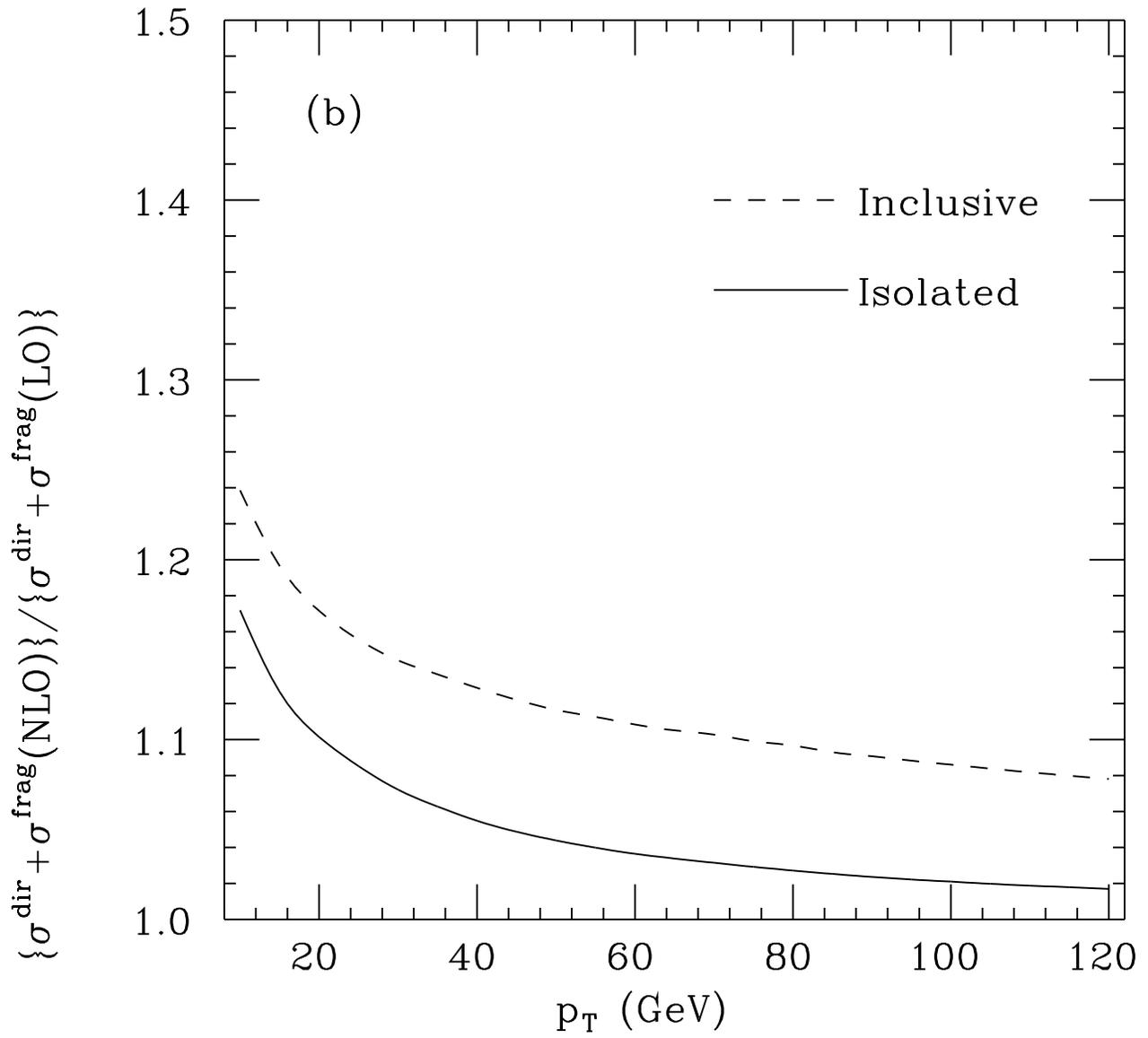
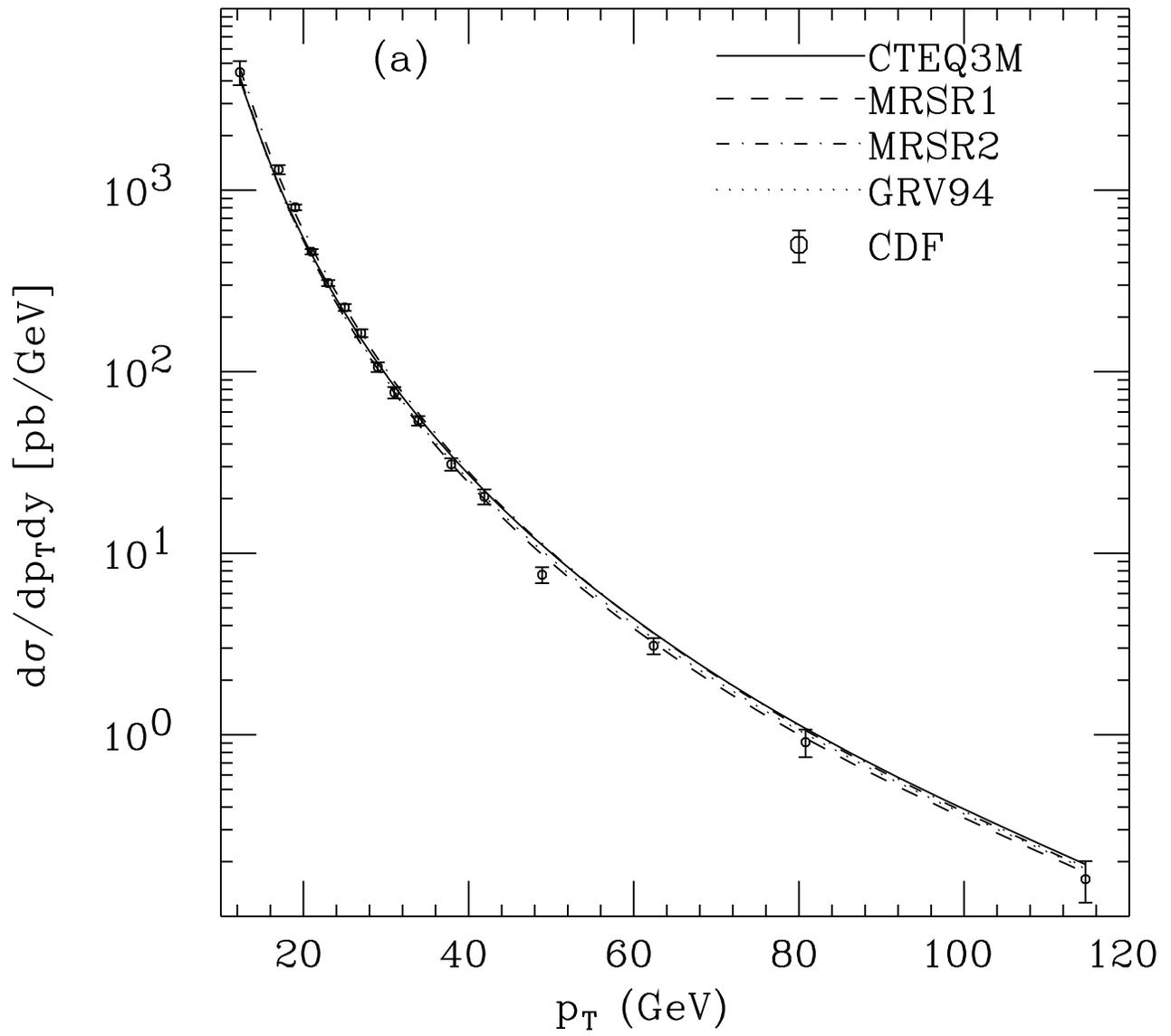
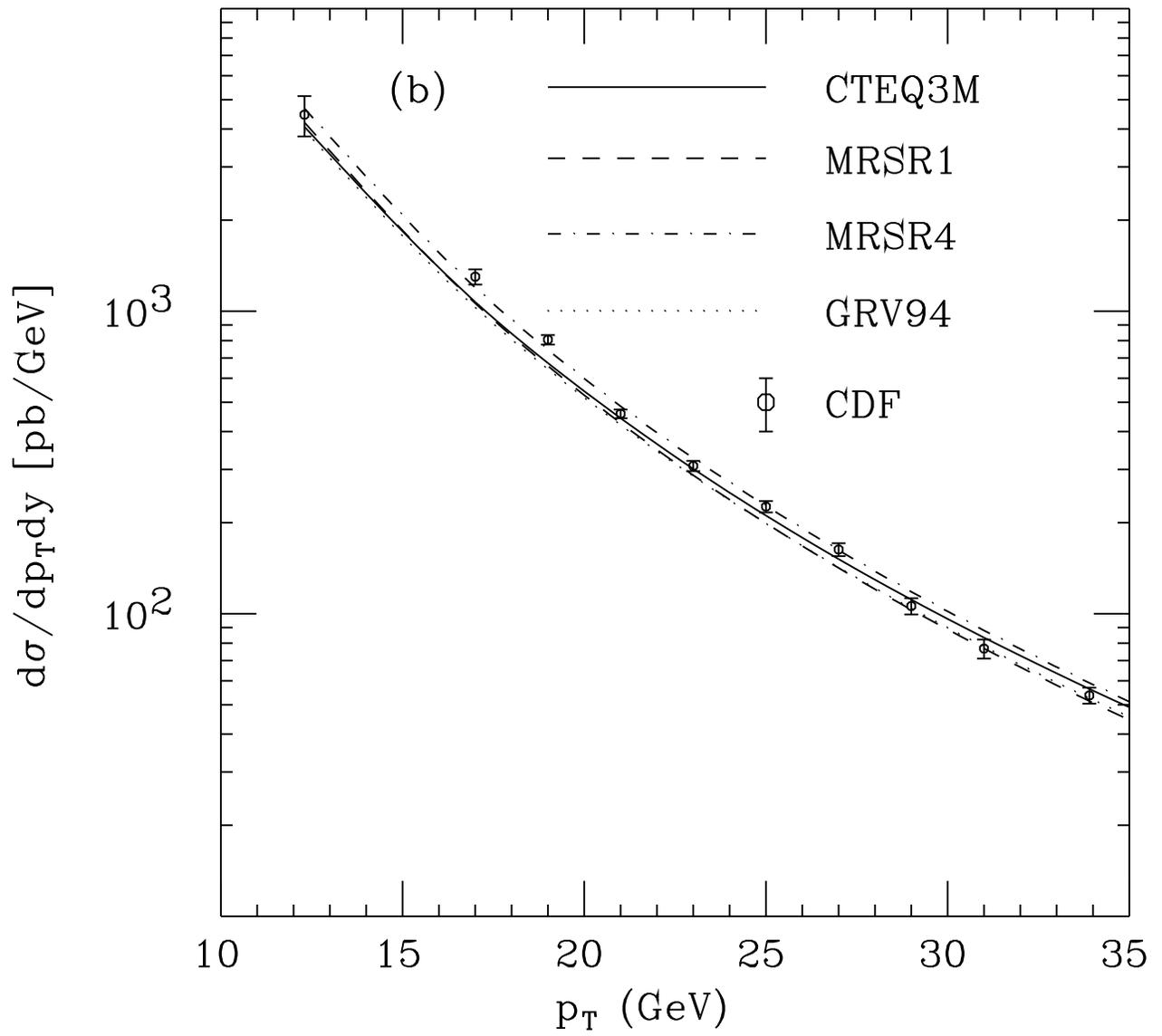
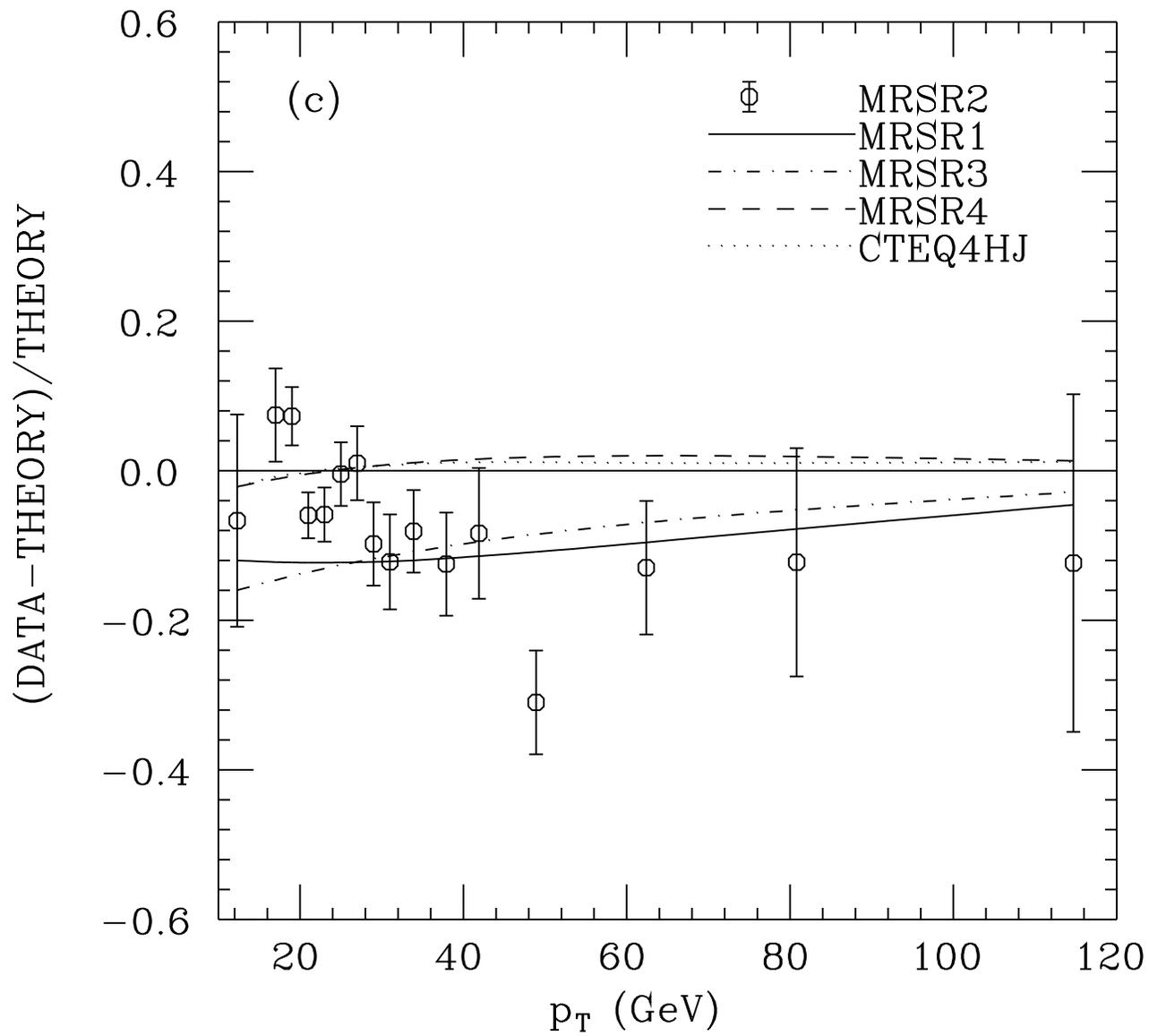
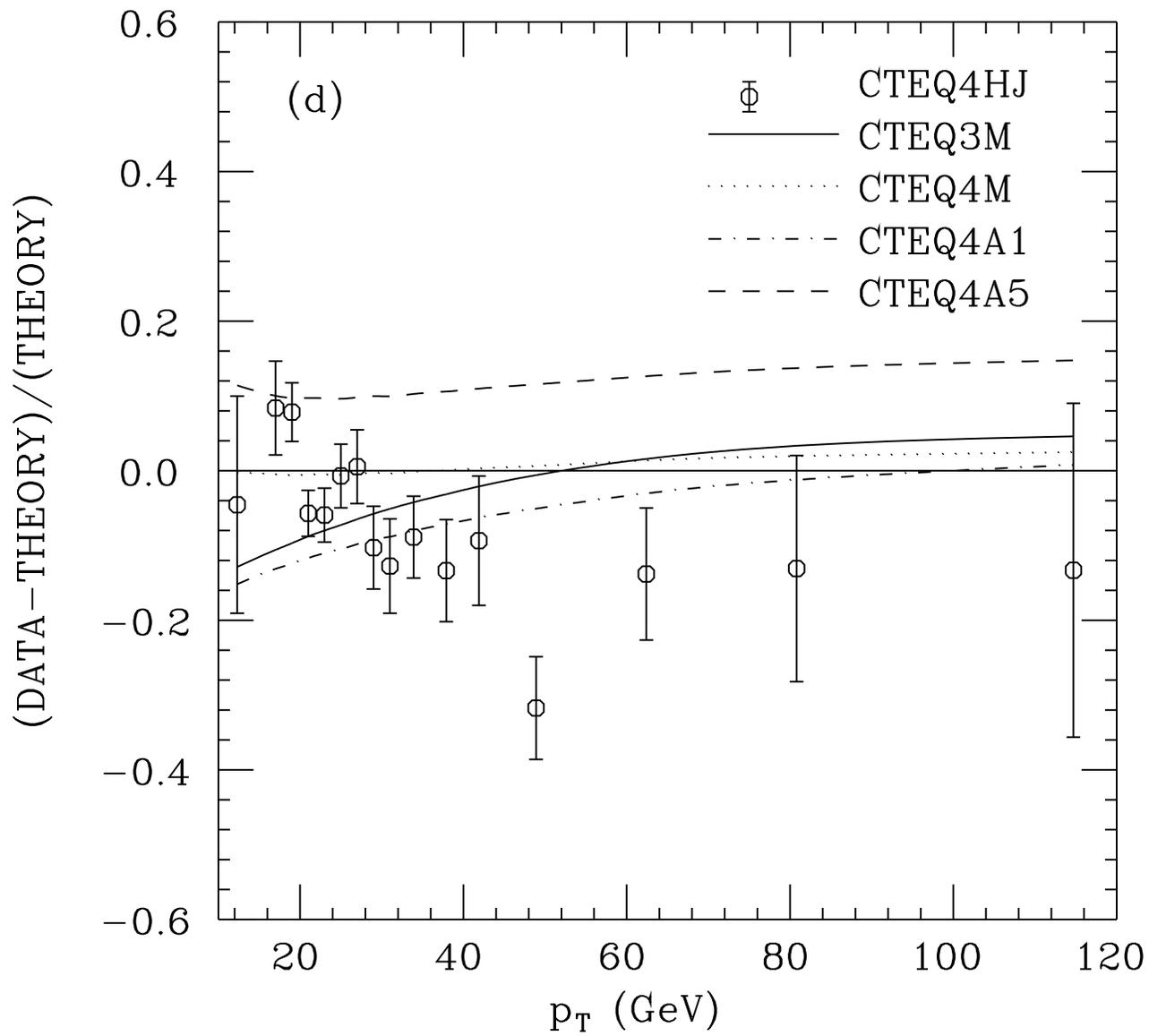
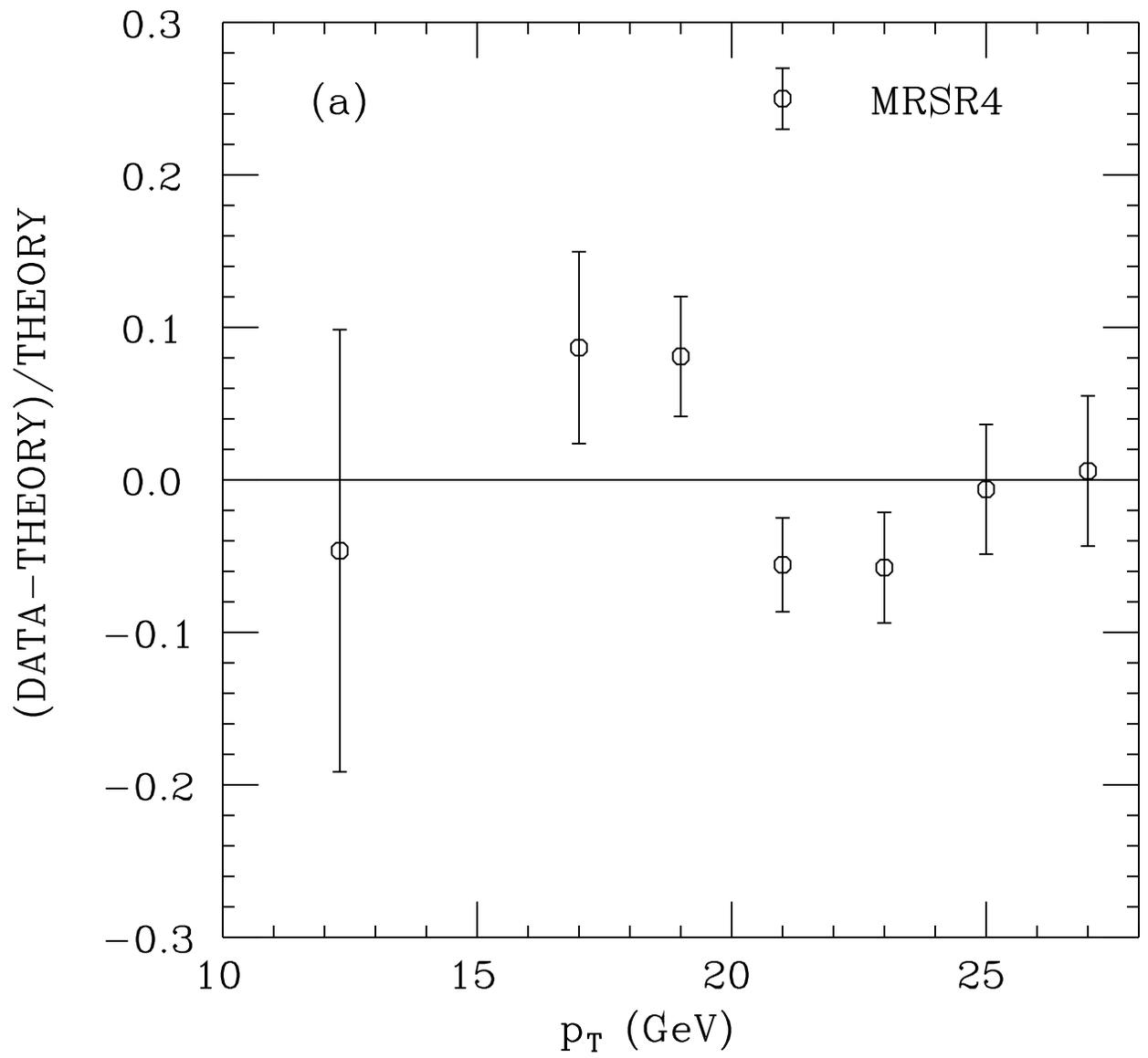
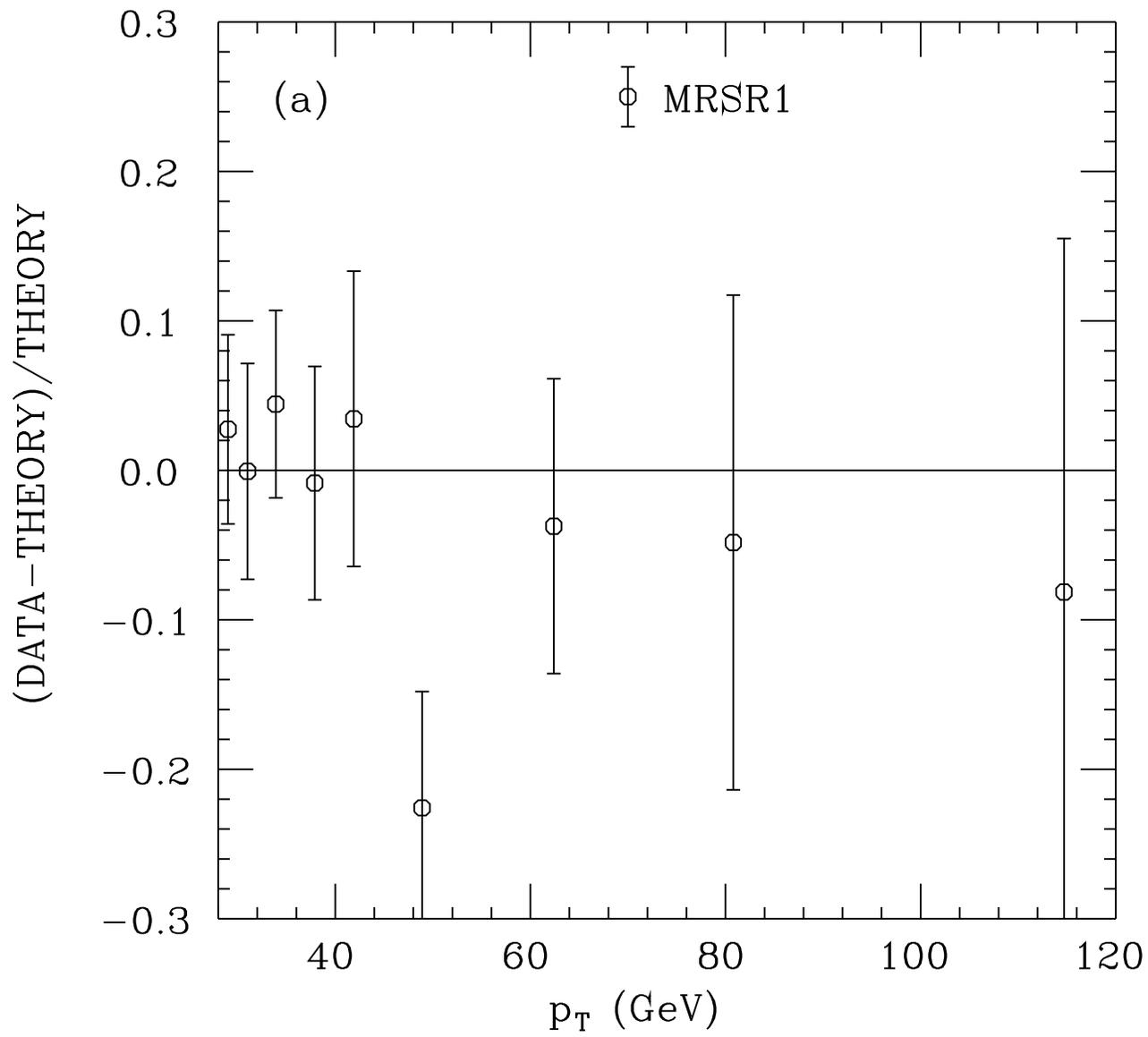
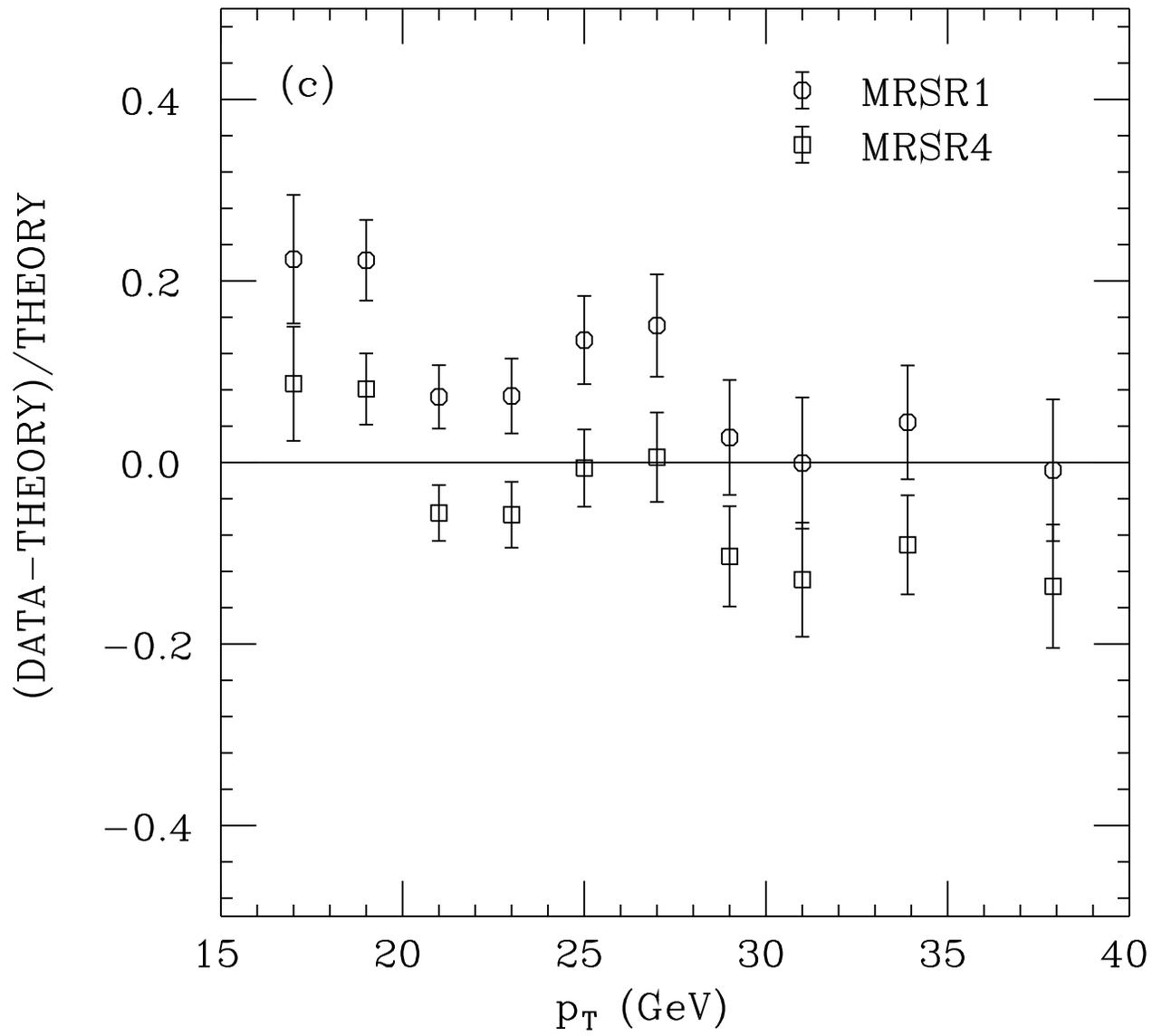
\end{document}